\documentclass[aps,amssymb,amsmath,
twocolumn,showpacs,superscriptaddress,
groupedaddress]{revtex4-1}

\usepackage{graphicx}
\usepackage{caption}
\usepackage{caption}
\usepackage[section]{placeins}
\usepackage{subfig}
\usepackage{dcolumn}
\usepackage{bm}
\usepackage[table,xcdraw]{xcolor}
\usepackage[flushleft]{threeparttable}
\usepackage{csquotes}
\usepackage{leftidx}
\usepackage{braket}
\usepackage{mathtools}
\usepackage{tikz}
\usepackage{epstopdf}
\usetikzlibrary{arrows.meta,calc,decorations.markings,math,arrows.meta,positioning}
\usepackage{graphicx}
\usepackage{hyperref}
\definecolor{OliveGreen}{cmyk}{0.64,0,0.95,0.40}
\definecolor{royalfuchsia}{rgb}{0.79, 0.17, 0.57}
\definecolor{vividauburn}{rgb}{0.58, 0.15, 0.14}

\begin{document}
\title{On the use of Slater-type spinor orbitals in Dirac-Hartree-Fock method.\\
Results for hydrogen-like atoms with super$-$critical nuclear charge.}

\author{A. Ba{\u g}c{\i}}
\email{ali.bagci@pau.edu.tr}
\affiliation{Department of Physics, Faculty of Arts and Sciences, Pamukkale University 20017 Denizli, Turkey}

\begin{abstract}
This work presents the formalism for evaluating molecular SCF equations, as adapted to four$-$component Dirac spinors, which in turn reduce to Slater$-$type orbitals with non$-$integer principal quantum numbers in the non$-$relativistic limit. The "catastrophe" which emerges for a charge numbers $Z>137$, in  solving  the  Dirac  equation  with a potential corresponding to a point$-$charge is avoided through using Slater$-$type spinor orbitals in the algebraic approximation. It is observed that, ground$-$state energy of hydrogen$-$like atoms reaches the negative$-$energy continuum $\left(-mc^2 \right)$ while critical nuclear charge $Z_{c}$, about $Z_{c}=160$. The difficulty associated with finding relations for molecular integrals over Slater$-$type spinors which are not$-$analytic in the sense of complex analysis at $r = 0$, is eliminated. Unique numerical accuracy is provided by solving the molecular integrals through Laplace expansion of Coulomb interaction and prolate spheroidal coordinates. New convergent series representation formulae are derived. The technique draws on previous work by the author and the general formalism is presented in this paper.
\begin{description}
\item[Keywords]
Dirac equations, Slater$-$type spinor orbitals, molecular integrals, analytical evaluation.
\item[PACS numbers]
... .
\end{description}
\end{abstract}
\maketitle

\section{\label{sec:intro}Introduction}
Methods developed on electronic structure calculations through the Schr{\"o}dinger equation have an almost definitive framework from the theoretical point of view. It is thus easy and advantageous to exactly specify the problem to be studied. For the Dirac equation on the other hand, no matter how specific the problem, a comprehensive approximation is absolutely necessary. This is such that a small improvement on that given problem may lead to a significant effect on whole theory.\\
This article is organised as follows: in the present introductory section the Dirac$-$Fock method and problems arising in relativistic calculations are defined, in general. The subjects of interest are introduced. In section \ref{sec:stso} the Slater$-$type spinor orbitals are described, suitable for solving the Dirac equation of hydrogen-like ultra-heavy atoms. Section \ref{sec:DHFR} gives algebraic Dirac$-$Fock formalism, which is general for atoms and molecules. Section \ref{sec:NonrelatMolInteg} gives the non-relativistic limit of overlap and two$-$electron integrals in molecules. Section \ref{sec:molauxrev} describes relativistic molecular auxiliary integrals useful in the Poisson equation solution of the Coulomb potential, contributing to Fock$-$Dirac matrix elements.

The problem of accounting for relativistic effects on molecules including heavy atoms is studied by analogous generalization of the independent particle model (Hartree$-$Fock approximation) \citep{1_Hartree_1928, 2_Fock_1930, 3_Fock_1978}. The Schr{\"o}dinger Hamiltonian is replaced by the Dirac Hamiltonian and the formalism is adapted to Quantum Electrodynamics (QED) \cite{4_Landau_1982, 5_Lindgren_2011}. The resulting equations are solved iteratively by writing them in form of generalized eigenvalue problem \citep{6_Ford_1974} via the linear combination of atomic spinors (LCAS) method \citep{7_Roothaan_1951, 8_Grant_1961, 9_Grant_1965, 10_Kim_1967, 11_Leclercq_1970, 12_Laaksonen_1988, 13_Quiney_1987, 14_Malli_1975, 15_Matsuoka_1980, 16_Pisani_1994, 17_Yanai_2001, 18_Quiney_2002, 19_Belpassi_2008} as follows:
\begin{align}\label{eq:1_DCHAMIL}
H_{DCB}=\sum_{i}h_{D}(i)+\sum_{i<j}V_{CB}\left(ij\right).
\end{align}
$H_{DCB}$ is the \textit{no$-$pair} relativistic Dirac$-$Coulomb$-$Breit many electron Hamiltonian \cite{20_Sucher_1980} in Born$-$Oppenheimer approximation and atomic units (a. u.). $h_{D}\left(i\right)$ is the one$-$electron Dirac operator for $i^{\text{\tiny th}}$ electron in a system,
\begin{align}\label{eq:2_ONEELECTDCHAMIL}
\hat{H}_{\mathit{D}}=c(\vec{\alpha}.\hat{\vec{p}})+\left(\beta-1\right)c^{2}-\frac{Z_{A}}{r_{Ai}},
\end{align}
\begin{align}\label{eq:3_ALPHA_UNIT_MAT}
\vec{\alpha}=\begin{pmatrix} 0& \vec{\sigma} \\ \vec{\sigma}& 0 \end{pmatrix}
\qquad\text{}\qquad
\beta=\begin{pmatrix} I& 0 \\ 0& {-I} \end{pmatrix},
\end{align}
where $\vec{\sigma}$ stands for Pauli spin matrices, $\hat{\vec{p}}$ is the momentum operator, $I$ is the $2\times2$ unit matrix and $c$ is the speed of light.
$Z_{A}$ is the nuclear charge of nucleus $A$, $r_{Ai}$ is the distance between nucleus $A$ and electron $i$. The second components in the Dirac$-$Coulomb$-$Breit Hamiltonian are the inter$-$electron Coulomb repulsion operator and frequency dependent Breit interaction, respectively. \\
Consider the Rayleigh quotient of the Dirac$-$Coulomb Hamilton $\left( \hat{H}_{\mathit{DC}} \right)$ operator for a closed$-$shell system. The wave$-$functions $\Psi$ is a single anti-symmetrized product of molecular spinors $\psi$, 
\begin{align}\label{eq:4_RELATSLATERDET}
\Psi
=\frac{1}{\sqrt{N!}}\sum_{p}\left(-1 \right)^{p}P\left[
\psi_{1}\left(\vec{r}_{1} \right) 
\psi_{2}\left(\vec{r}_{2} \right)
...
\psi_{N}\left(\vec{r}_{N} \right)
\right],
\end{align}
here, $P$ permutation operator. The $\psi$ are expanded by the LCAS method in terms of atomic spinors,
\begin{align}\label{eq:5_LCASSTSO}
\psi_{p}
=\sum_{q}^{N} X_{q}C_{pq},
\end{align}
in matrix form is,
\begin{align}\label{eq:6_VariationDCHAMIL}
R\left(\Psi \right)=
\frac{\Braket{ \Psi |\hat{H}_{\mathit{DC}}  | \Psi}}{\Braket{ \Psi | \Psi}}.
\end{align}
The $\psi_{p}$ are taken to be orthonormal; that is,
\begin{align}\label{eq:7_MOLSPINORORTH}
\int_{}^{}\psi_{p}^{\dagger}\psi_{q}dV
=\delta_{pq},
\end{align}
where summation runs over the molecular spinors. The matrix form of the Hartree$-$Fock self$-$consistent field equations is given in Eqs. (\ref{eq:4_RELATSLATERDET}-\ref{eq:7_MOLSPINORORTH}),  \cite{7_Roothaan_1951,10_Kim_1967,21_Grant_2007},
\begin{align}\label{eq:8_DHFRFOURCOMPACT}
FC=SCE
\end{align}
in terms of matrix elements, we obtain:
\begin{align}\label{eq:9_DHFRFOUREXP1}
\sum_{q=1}^{N}F_{pq}C_{iq}=\epsilon_{i}\sum_{q=1}^{N}S_{pq}C_{iq},
\end{align}
with, $\epsilon_{i}$ is the orbital energy of the $i^{\text{\tiny th}}$ molecular spinor, $F_{pq}$ represent the elements of the relativistic Dirac$-$Fock matrix. \\
The atomic spinors are the four$-$component vectors \cite{22_Dirac_1930, 21_Grant_2007} whose components are the scalar wave$-$functions,
\begin{align}\label{eq:10_FOURCWAVEFUNC}
X=\begin{pmatrix} \chi^{\beta}_{1}\vspace{1.5 mm}\\ \chi^{\beta}_{2}\vspace{1.5 mm}\\ \chi^{-\beta}_{3}\vspace{1.5 mm}\\ \chi^{-\beta}_{4} \end{pmatrix},
\end{align} 
with $\beta= \pm 1$. The preferred nomenclature for the positive energy solutions, for the upper two$-$ and the lower two$-$components of atomic spinors are \textit{large} $\left( L \right)$ and \textit{small} $\left( S \right)$ components, respectively \cite{23_Foldy_1950}. The lower components go to zero in the non$-$relativistic limit and the upper components thus become a solution of the corresponding non$-$relativistic equation, i.e. the Schr{\"o}dinger equation. The spectrum obtained from the solution is the complete set of positive$-$ and negative$-$energy continuum states together with the discrete spectrum of bound states \cite{21_Grant_2007, 22_Dirac_1930, 24_Greiner_2000}.  Note that, representation of the whole spectrum is needed. The contribution of negative energy continuum states can significantly improve accuracy of solutions \cite{25_Grant_2010}. This makes the Hartree$-$Fock approximation suitable not only for studying the relativistic many$-$body perturbation theory via the linear combination of atomic spinor (LCAS) method \citep{8_Grant_1961, 9_Grant_1965, 10_Kim_1967, 11_Leclercq_1970, 12_Laaksonen_1988, 13_Quiney_1987, 14_Malli_1975, 15_Matsuoka_1980, 16_Pisani_1994, 17_Yanai_2001, 18_Quiney_2002, 19_Belpassi_2008, 26_Grant_1980, 27_Ishikawa_1991, 28_Ishikawa_1993, 29_Koc_1994, 30_Ishikawa_1997, 31_Ishikawa_2001, 32_Quiney_1997, 33_Quiney_1998, 34_Quiney_2004, 35_Saue_1996, 36_Saue_1997, 37_Saue_2011, 38_Pyykko_2012, 39_Motoumba_2019, 40_Si_2018, 41_Indelicato_1995} but also the quantum electrodynamics (QED) effects \cite{10_Kim_1967, 13_Quiney_1987, 14_Malli_1975, 25_Grant_2010, 32_Quiney_1997, 33_Quiney_1998, 34_Quiney_2004, 37_Saue_2011, 40_Si_2018, 42_Kutzelnigg_2012, 43_Fleig_2012}.

The LCAS method, however, is based on minimization (according to the variation principle). It only works rigorously if the spectrum has a lower bound. The unbounded property of the spectrum obtained from the solution of the Eq. (\ref{eq:9_DHFRFOUREXP1}) on the other hand, may cause \textit{variational collapse} \cite{44_Schwarz_1982, 45_Schwarz_1982} or appearance of {\sl spurious} un$-$physical states, between the lowest bound state and negative energy continuum \cite{46_Drake_1981, 47_Goldman_1985}. It is overcome by choosing atomic spinors satisfying the \textit{kinetic$-$balance} condition \cite{48_Lee_1982, 49_Stanton_1984, 50_Dyall_2012},
\begin{align}\label{eq:11_KINBALRELAT}
\lim_{c \to \infty} c\chi^{\vert \beta \vert}=\frac{1}{2m_{0}}\vec{\sigma}.\hat{\vec{p}}\lim_{c \to \infty}\chi^{\vert \beta \vert},
\end{align}
which also ensures that the non$-$relativistic limit is correct and the spectrum is separated into positive and negative energy parts.

The aim of this research in general, is to investigate limits of the solution for the Dirac equation while the {\bf point$-$like model} is considered for nucleus. It is obvious from exact solution of the Dirac equation for the Coulomb potential \cite{22_Dirac_1930} that for atoms with nuclear charge larger than $Z=137$ the electron collapses to the center, i.e., an atomic nucleus with charge $Z>137$ does not exist in nature. This inference for the Coulomb potential seems to be void in the algebraic solution if Slater$-$type spinor orbitals \cite{51_Bagci_2016} as basis sets are used. These basis functions also pave a way to overcome difficulties arise in evaluation of molecular integrals, constitute the matrix elements Dirac$-$Fock equations. 

The four$-$component formalism for relativistic SCF equations will now be revisited, accordingly.
\begin{figure}[htp!]
\centering
\begin{tikzpicture}[scale=0.75]
\coordinate (Origin) at (0,0,0);
\draw[thick,->] (0,0,0) -- (7,0,0) node[anchor=north east]{$y$};
\draw[thick,->] (0,0,0) -- (0,7,0) node[anchor=north west]{$z$};
\draw[thick,->] (0,0,0) -- (0,0,7) node[anchor=south]{$x$};

\coordinate (A) at (2,5,2);
\coordinate (B) at (7,4,2);
\coordinate (C) at (2.5,2.7,2.5);
\coordinate (D) at (6,2.5,2);
\coordinate (E1) at (2.5,7,1);
\coordinate (E2) at (7.5,7,3);

\draw [fill=blue] (Origin) circle  node [left] {O};
\draw [fill=black] (A) circle (3pt) node [below left=-0.3cm and 0.00 of A] {a};
\draw [fill=black] (B) circle (3pt) node [right] {b};
\draw [fill=black] (C) circle (3pt) node [below=0.1cm and 0.00 of C] {c};
\draw [fill=black] (D) circle (3pt) node [below=0.1cm and 0.00 of D] {d};
\draw [fill=red] (E1) circle (2pt) node [above] {$1$};
\draw [fill=red] (E2) circle (2pt) node [above] {$2$};

\draw[black,thick] (2,5,2) -- (7,4,2);
\draw[black,thick] (2,5,2) -- (2.5,2.7,2.5);
\draw[black,thick] (2.5,2.7,2.5) -- (6,2.5,2);
\draw[black,thick] (7,4,2) -- (6,2.5,2);

\draw[black,thick] (0,0,0) -- (2,5,2) node [black, pos=0.6, left] {$\vec{R}_{a}$};
\draw[black,thick] (0,0,0) -- (7,4,2)node [black,pos=0.65, above] {$\vec{R}_{b}$};
\draw[black,thick] (0,0,0) -- (2.5,2.7,2.5) node [black,pos=0.85, left] {$\vec{R}_{c}$};
\draw[black,thick] (0,0,0) -- (6,2.5,2) node [black,pos=0.75, below] {$\vec{R}_{d}$};

\draw[black,dashed,line width=1.0pt] (2,5,2) -- (2.5,7,1) node [black,pos=0.6, left]{};
\draw[black,dashed,line width=1.0pt] (7,4,2) -- (2.5,7,1) node [black,pos=0.85, right]{};
\draw[black,dashed,line width=1.0pt] (2.5,2.7,2.5) -- (2.5,7,1) node [black,pos=0.7, right]{};
\draw[black,dashed,line width=1.0pt] (6,2.5,2) -- (2.5,7,1) node [black,pos=0.72, right]{};

\draw[black,dashed,line width=0.5pt] (2,5,2) -- (7.5,7,3) node [black,left=0.1cm and 0.5cm of E2]{};
\draw[black,dashed,line width=0.5pt] (7,4,2) -- (7.5,7,3) node [black,pos=0.69, right]{};
\draw[black,dashed,line width=0.5pt] (2.5,2.7,2.5) -- (7.5,7,3) node [black,pos=0.80, left]{};
\draw[black,dashed,line width=0.5pt] (6,2.5,2) -- (7.5,7,3) node [black,pos=0.65, left]{};

\draw[black,line width=0.5pt] (2.5,7,1) -- (7.5,7,3) node [black,pos=0.55, above] {$\vec{r}_{12}$};

\coordinate (OriginA) at (2,5,2);
\coordinate (OriginB) at (7,4,2);
\coordinate (OriginC) at (2.5,2.7,2.5);

\end{tikzpicture}
\caption {\label{fig:GeometryFig} Depiction of the coordinates for motion of electrons in the field of four stationary Coulomb centers, namely $a$, $b$, $c$, $d$, where $a=\left\lbrace z_{a}, y_{a},x_{a} \right\rbrace$, $b=\left\lbrace z_{b},y_{b},x_{b} \right\rbrace$, $c=\left\lbrace z_{c},y_{c},x_{c} \right\rbrace$, $d=\left\lbrace z_{d},y_{d},x_{d} \right\rbrace$, $\left\lbrace z, y, x \right\rbrace$ are the axes of Cartesian coordinates.}
\end{figure}
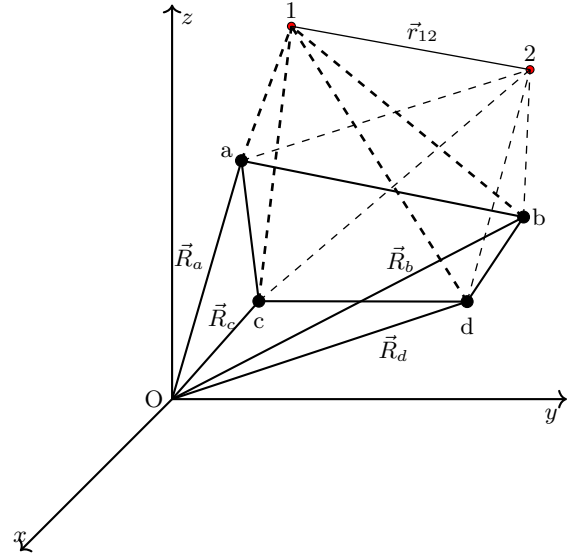
\section{\label{sec:stso}Slater$-$type atomic spinors for relativistic calculations of heavy and super$-$heavy elements}
The Slater$-$type spinor orbitals (STSOs) as atomic spinors have the functional form of node$-$less $L-$spinors \cite{21_Grant_2007}, or those with the fewest nodes, characterized by minimum values of radial quantum numbers. They are are advantageous to use in the LCAS method. The STSOs can be considered as relativistic analogues of Slater$-$type functions with non$-$integer principal quantum numbers. The STSOs are given as:
\begin{align}\label{eq:12_STSO}
X_{nljm}\left(\zeta, \vec{r} \right)
=\begin{pmatrix}
\chi_{nljm}^{\beta 0}\left(\zeta, \vec{r}\right)
\vspace{1.5 mm}\\
\chi_{nljm}^{\beta 1}\left(\zeta, \vec{r}\right)
\vspace{1.5 mm}\\
\chi_{nljm}^{-\beta 0}\left(\zeta, \vec{r}\right)
\vspace{1.5 mm}\\
\chi_{nljm}^{-\beta 1}\left(\zeta, \vec{r}\right)
\end{pmatrix},
\end{align}
here:
\begin{align} \label{eq:13_STSOF}
\chi_{nljm}^{\beta \varepsilon}\left(\zeta, \vec{r}\right)
=f^{\beta}_{nlj}(\zeta,r) \Omega_{ljm}^{\beta \varepsilon} \left(\theta, \vartheta\right),
\end{align}
\begin{align} \label{eq:14_RADIALSTSO}
f^{\beta}_{nlj}(\zeta,r)
=\left\{{A_{nlj}^{\beta}r^{n}+\zeta B_{nlj}^{\beta}r^{n+1}}\right\}e^{-\zeta r},
\end{align}
here, $\beta$ represents large$-$ and small$-$components of STSOs, $\left\lbrace n,l,j,m \right\rbrace$ are the principal, angular, total angular and secondary total angular momentum quantum numbers with $n \in \mathbb{R}^{+}$, $0\leq l \leq \lfloor n \rfloor-1$, $j=l\mp 1/2$, $-j \leq m \leq j$ and $\lfloor n \rfloor$ stands for the integer part of $n$ respectively. $\zeta$ are orbital parameters. Note that formalism symmetry, with two$-$radial components is provided by this representation.\\
The $\Omega_{ljm}^{\beta \varepsilon}$ are the spin $\frac{1}{2}$ spinor spherical harmonics \cite{52_Davydov_1976},
\begin{align}\label{eq:15_SSHF}
\Omega_{ljm}^{\beta \varepsilon}\left(\theta, \vartheta\right)
=\text{\large $a$}_{l_{\beta}jm \left( \varepsilon \right)}
\text{\large $\eta$}_{m\left( \varepsilon \right)}
Y_{l_{\beta} {m\left( \varepsilon \right)}}\left(\theta, \vartheta\right),
\end{align}
where, the values of $l_{\beta}$ are determined by $l_{\beta}=j+\frac{\beta}{2}$, $\varepsilon$ stands to represent spherical part of each component of STSOs and $\eta_{m\left(\varepsilon \right)}=\mathtt{i}^{\vert m\left(\varepsilon \right)\vert-m\left(\varepsilon \right)}$. The quantities $\text{\large $a$}$ are the Clebsch$-$Gordan coefficients. They are given through Wigner$-$3j symbols \citep{53_Wigner_1959, 54_Varshalovich_1988, 55_Wei_1999} as,
\begin{multline}\label{eq:16_CGCOEFF}
\text{\large $a$}_{ljm\left( \varepsilon \right)}
=\left(l\frac{1}{2}m\left( \varepsilon \right) \frac{1}{2}-\varepsilon
\bigg\vert l\frac{1}{2}jm\right)\\
=\frac{\left(-1\right)^{\frac{1}{2}-l-m}}{\sqrt{2j+1}}
\begin{pmatrix}
  l & \frac{1}{2} & j \\
  m\left(\varepsilon \right) & \frac{1}{2}-\varepsilon & -m
 \end{pmatrix}.
\end{multline}
$Y_{lm_{l}}$ are the complex spherical harmonics $(Y^{*}_{lm_{l}}=Y_{l-m_{l}})$,
\begin{align} \label{eq:17_SPHERICALHARM}
	Y_{l\vert m_{l} \vert}(\theta, \vartheta)
	=\frac{1}{2\pi}\mathcal{P}_{l|m_{l}|}\left(cos\theta \right)e^{\mathtt{i}m_{l}\vartheta}.
\end{align}
It differs from the Condon$-$Shortley phase by a sign factor $(-1)^{m_{l}}$ \cite{56_Condon_1970}. $\mathcal{P}_{lm_{l}}(x)$ is the associated Legendre function, $m_{l} \equiv m\left(\varepsilon \right)$ are the magnetic quantum numbers, respectively.\\
Radial parts of STSOs satisfy the proper symmetry and functional relationship between large$-$ and small$-$components for any values of $n$ as follows:
\begin{multline} \label{eq:18_TCFSTSODIFEQ}
\frac{\partial}{\partial r}f_{n\kappa}^{\beta}\left(\zeta, r\right)
=-\beta\frac{\kappa}{r}f_{n\kappa}^{\beta}\left(\zeta, r\right) \\
+\left(\frac{\beta N_{n\kappa}-n-\delta_{\vert \kappa\vert \kappa}}{r}+ \zeta \right)f_{n\kappa}^{-\beta}\left(\zeta, r\right).
\end{multline}
They also obey both the cusp condition at the nucleus \cite{57_Kato_1957} and exponential decay at long range \cite{58_Agmon_1985}. The assertion that, using point$-$like nucleus model causes a \textit{weak singulariy} at the origin \cite{59_Singulariy_2017} i.e., the pair of radial functions do not fulfill the conditions,
\begin{align}\label{eq:19_WEAKSING}
\begin{pmatrix}
f_{nlj}^{\beta}\left(\zeta, 0\right)
\vspace{1.5 mm}\\
f_{nlj}^{-\beta}\left(\zeta, 0\right)
\end{pmatrix}
=
\begin{pmatrix}
0
\\
0
\end{pmatrix},
\hspace{3.3mm}
\lim_{r\rightarrow \infty}
\begin{pmatrix}
f_{nlj}^{\beta}\left(\zeta, r\right)
\vspace{1.5 mm}\\
f_{nlj}^{-\beta}\left(\zeta, r\right)
\end{pmatrix}
=
\begin{pmatrix}
0
\\
0
\end{pmatrix}
\end{align}
is, therefore refuted and disadvantages of using Slater$-$type radial in atomic spinors may be overcome since, $\gamma \equiv n$ and $n$ can have values that $n=\vert \kappa \vert$ which is also independent from speed of light. The STSOs are also of the same form as $S-$spinors \cite{21_Grant_2007} if
\begin{align*}
n=\gamma=\sqrt{\kappa^2-\left(\alpha Z\right)^2}
\end{align*}
except that their radial parts are coupled for large$-$ and small$-$components. They satisfy the criteria summarized by Grant \cite{21_Grant_2007} for constructing a relativistic basis set for radial amplitudes:
\begin{itemize}
\item[1.]{The Dirac Hamiltonian imposes functional relations between the upper and lower components which must be respected.}
\item[2.]{Care must be taken to ensure functions have the correct asymptotic form near the nuclear Coulomb singularity.}
\item[3.]{The relativistic equations must reproduce the non$-$relativistic equivalents asymptotically as $c\rightarrow \infty$.}
\item[4.]{If possible, the basis sets should be complete in a suitable Hilbert space so that (theoretical) convergence as the basis set enlarge can be guaranteed.}
\end{itemize}

The restriction $\alpha Z <1$ in the point-like model of nucleus \cite{59_Singulariy_2017} (see also references therein) therefore no longer applies. Here, $\alpha$ is the fine structure constant.
As it is stated in our previous work \cite{51_Bagci_2016}, this facilitates studying new advances in atomic, molecular, and nuclear physics such as laser$-$matter interaction \cite{60_Mourou_2006}, electrons have been subjected to a very intense magnetic field \cite{61_Selsto_2009} also the also exotic atoms which are very sensitive to quantum electrodynamic effects \cite{62_Pohl_2013}. The hydrogen$-$like muonium atom $\left(\mu^{+}e^{-}\right)$, which consists of two point$-$like leptons of different types. It is obtained by replacing the hadronic nucleus (proton) in a hydrogen atom with the positive muon $\left( \mu^{+} \right)$. Absence of any hadronic constituent leads to energy levels to be calculated {\it in fine}. It is an ideal object for testing quantum electrodynamics and the behavior of the muon as a point$-$like heavy leptonic particle \cite{63_Jungmann_1992}.

The primary objective of the present paper is to study the usefulness of STSOs. Accordingly:
\begin{itemize}
\item{The results obtained for hydrogen$-$like atoms in previous paper are improved through increasing the value of upper milit of summation in LCAS.}
\end{itemize}
One of the important features of the hydrogen atom Dirac$-$Hamiltonian is that the bound state energy levels form a super$-$symmetric pattern. They appear as functions of $\kappa^2$ and the radial quantum number $n_{r}$, $n_{r}=n-\vert \kappa \vert$. They are separated according to the value of $\vert \kappa \vert=j+\frac{1}{2}$. And the degeneracy of an energy level is $2j+1$ \cite{24_Greiner_2000}.
\begin{itemize}
\item{Convergence of the degenerate excited states of hydrogen$-$like atoms for a specific value of orbital parameter are investigated additionally, where the principal quantum numbers are chosen such that $n=\gamma^{-}=\sqrt{\kappa^2-{Z^2}/{c^2}}$, $n \in \mathbb{Z}^{+}\left(n=\vert \kappa \vert \right)$, $n=\gamma^{+}=\sqrt{\kappa^2+{Z^2}/{c^2}}$.}
\item{The ground and some excited state energy eigenvalues are presented depending on the values of nuclear charge, where $110 \leq Z \leq 160 \hspace{2mm} a.u.$.}
\end{itemize}

A power function such as $z^a=e^{a\log{z}}$ is analytic at $z_{0}=0$ if $a \in \mathbb{Z}$ is an integer \citep{64_Olver_2018}. This implies that, expanding the power function near the origin by a power series only converges if $a$ is non$-$negative integer.
\begin{align}\label{eq:20_POWFONKSER}
	f_{p}(z)=\sum_{i=0}^{\infty} w_{i}(z-z_{0})^{i},
\end{align}
where, $z_{0}$ is a constant, and $z$ varies around $z_{0}$, $w_{i}$ represents the coefficient of the i$^{th}$ term; they essentially correspond to the derivatives of $f_{p}$ at $z_{0}$. The exponent $n$ of power function $r^n$ occurring in Eq. (\ref{eq:14_RADIALSTSO}), on the other hand, is in set of positive real numbers ($n \in \mathbb{R}^{+}$). Power series representation of $r^n$ for finite values of upper limit of summation is semi$-$convergent \cite{65_Weniger_2008, 66_Weniger_2012}.\\
One of the main advantages of using Slater$-$type spinors in relativistic molecular electronic structure calculations is they avoid the above difficulty. The author in his previous papers \cite{67_Bagci_2014, 68_Bagci_2015, 69_Bagci_2015} avoided such difficulty through using numerical methods, namely, global adaptive method with Gauss$-$Kronrod numerical integration extension. Evaluation of the relativistic molecular integrals problem was solved regarding accuracy via the \textit{Mathematica} programming language \cite{70_MathematicaProg}. The \textit{Mathematica} programming language is, however, suitable only for bench$-$marking in the view of calculation times. Necessity of deriving analytical relations thus, obvious not only for mathematical consistency but also applications. This task has been accomplished by the author \cite{71_Bagci_2018_1, 72_Bagci_2018_2} using the formulae given in previous unpublished versions of the present paper \cite {73_Bagci_2017}. The analytical formulae given here, through series representation of incomplete beta functions and in terms of integrals involving Appell functions also reduced to
series representation formulae for incomplete beta functions.

The second objective of the present work is to derive analytical formulae for calculating the relativistic molecular integrals. Here, the sub$-$functions at the summations are calculated numerically in order to prove convergence of series representation.
\begin{itemize}
\item{Convergent series representation formulae, which are suitable to be written in in any high$-$level programming language such as \textit{FORTRAN} or \textit{C++}, for two-center two-electron molecular integrals are derived. The results obtained are compared with the given benchmark values in the previous papers.}
\end{itemize}
\section{\label{sec:DHFR}The Dirac$-$Hartree$-$Fock Equations in Algebraic Approximation}
Solution of the Dirac equation for many$-$electron systems via the algebraic approximation through Eq. (\ref{eq:9_DHFRFOUREXP1}) are mainly based on two approaches. These approaches are classified by representation of spinors in which direct use of the Eq. (\ref{eq:10_FOURCWAVEFUNC}) in explicit form of Dirac-Fock equations is referred to as {\sl four$-$component spinor approach} (Dirac picture) \cite{22_Dirac_1930}. Representing the Dirac equation in two$-$component form utilizing from Foldy-Wouthuysen transformation \cite{23_Foldy_1950} and extending the problem to many$-$particle case is referred to as {\sl two$-$component spinor approach} (Newton-Wigner picture) \cite{74_Thaller_1992, 75_Autschbach_2000}. Current studies require representation of both positive and negative energy branches of spectrum since two$-$component calculations are beyond relativistic treatment of the atomic or molecular electronic structure but required in capturing most electron correlation at the relativistic level \cite{25_Grant_2010, 76_Schwerdtfeger_2015}. The complete picture of the spectrum is obtained from solution of four$-$component form of the Dirac equation and clear separation between positive and negative energy branches is seen as essential prerequisite \cite{34_Quiney_2004}. In addition to proper choice of basis function this require avoiding {\sl continuum dissolution} \cite{77_Brown_1951} arising from constructing the many$-$electron Hamiltonian with a relativistic one$-$electron part and non$-$relativistic two$-$electron term. The bound state and the continuum spectra are coupled by electron$-$electron interaction. By following the steps clearly outlined in \cite{20_Sucher_1980, 50_Dyall_2012} this difficulty is eliminated.

Several four$-$component {\sl ab$-$initio} atomic and molecular programs such as GRASP \cite{78_Dyall_1989}, MOLFDIR \cite{79_Visscher_1994}, DIRAC \cite{80_DIRAC_2018}, BERTA \cite{33_Quiney_1998, 81_Grant_2000} and quite recently BAGEL \cite{82_Shiozaki_2018} have been developed. GRASP uses point nuclei and is coded for atomic calculations. All the other software considers finite$-$sized nuclei. This results from the absence of methods to calculate the constituent matrix elements in the algebraic approximation for the point$-$like model of nucleus. It is imperative in this case to use the exponential$-$type spinor orbitals and was previously assumed that they do not fulfill the conditions required for relativistic calculations of transactinide elements (super$-$heavy elements). The relativistic effect for these elements are approximately $\sim\left(Z\alpha\right)^2$ or larger. They are not naturally found on Earth. They have to be synthesized by nuclear fusion reaction with heavy ion particles \cite{83_Hofmann_2011, 84_Grainer_2015}. Possibility for synthesis of super$-$heavy nuclei up to a nuclear charge $Z=122$ has been revealed in recent studies \cite{85_Oganessian_2011, 86_Manjunatha_2019} and the discussion on feasibility of such chemical experiments for higher nuclear charges is continuing intensively. The main difficulty experimentally results from short half-life of heavy nuclei. All beyond nuclear charge $Z=82$ are radioactive. Beyond nuclear charge $Z=104$, the half-life is too short that practical difficulty of collecting a sample is critical. Design of such difficult experiments relies on predetermined knowledge of the electronic structure and chemical behavior of the these super$-$heavy elements. For this, one requires accurate relativistic electronic structure calculations \cite{76_Schwerdtfeger_2015}.

Continuing to discuss the point$-$like model of nucleus in this context, it may be said that the mathematical difficulties mentioned above may no longer valid. The first point to highlight is that the exponent of radial amplitudes of the STSOs can have values such that $n=\vert \kappa \vert$ and the relativistic molecular integrals are easily be represented in terms of known non$-$relativsitic molecular integrals over Slater$-$type orbitals with integer principal quantum numbers \cite{87_Slater_1930},
\begin{align} \label{eq:21_NSTOs}
\chi_{nlm_{l}} \left(\zeta, \vec{r}\right)=
\frac{\left(2\zeta \right)^{n+1/2}}{\sqrt{\left((2n\right)!}}r^{n-1}e^{-\zeta r}Y_{lm_{l}}(\theta,\vartheta),
\end{align}
here, $m_{l}$ is the magnetic quantum number. Consider the three$-$ and four$-$center integrals. They must be represented in terms of the analytically expressed two$-$center molecular integrals. The translation methods which are used to express a single Slater$-$type orbital placed at a certain point of space as a series expansion involving quantities located at a different center \cite{88_Guseinov_1985} are still available.
\begin{multline}\label{eq:22_TransSTOs}
\chi_{nlm_{l}} \left(\zeta, \vec{r}_{a}\right) \\
=\sum_{n'l'm'}V_{nlm_{l},n'l'm'_{l}}\left(\zeta, \vec{R}_{ab}\right)
\chi_{n'l'm'_{l}} \left(\zeta, \vec{r}_{b}\right)
\end{multline}
where, $V$ are the expansion coefficients.  The expansion coefficients are usually represented in terms of two$-$center overlap integrals, are defined in the following section.\\
The second is that, while $n=\gamma$ methods for evaluation of molecular integrals up to a three$-$center have already been developed in both numerical and analytical approaches.

By Briefly revisiting explicit form of four$-$component Dirac$-$Fock formalism of the Dirac$-$Coulomb Hamiltonian regarding the constitute matrix elements and considering the relativistic spinors basis, the notation used in this paper the Eq. (\ref{eq:9_DHFRFOUREXP1}) is written as,
\onecolumngrid
\begin{multline}\label{eq:23_DHFRFOUREXP}
\begin{pmatrix}
f_{pq}^{\beta\varepsilon \beta\varepsilon} & f_{pq}^{\beta\varepsilon \beta(\varepsilon+1)} & f_{pq}^{\beta\varepsilon -\beta\varepsilon} & f_{pq}^{\beta\varepsilon -\beta(\varepsilon+1)}\vspace{2mm}
\\
f_{pq}^{\beta(\varepsilon+1) \beta\varepsilon} & f_{pq}^{\beta(\varepsilon+1) \beta(\varepsilon+1)} & f_{pq}^{\beta(\varepsilon+1) -\beta\varepsilon} & f_{pq}^{\beta(\varepsilon+1) -\beta(\varepsilon+1)}\vspace{2mm} 
\\
f_{pq}^{-\beta\varepsilon \beta\varepsilon} & f_{pq}^{-\beta\varepsilon \beta(\varepsilon+1)} & f_{pq}^{-\beta\varepsilon -\beta\varepsilon} & f_{pq}^{-\beta\varepsilon -\beta(\varepsilon+1)}\vspace{2mm}  
\\
f_{pq}^{-\beta(\varepsilon+1) \beta\varepsilon} & f_{pq}^{-\beta(\varepsilon+1) \beta(\varepsilon+1)} & f_{pq}^{-\beta(\varepsilon+1) -\beta\varepsilon} & f_{pq}^{-\beta(\varepsilon+1) -\beta(\varepsilon+1)}
\end{pmatrix}
\begin{pmatrix*}[c]
c_{pq}^{\beta\varepsilon}\vspace{2mm}
\\
c_{pq}^{\beta(\varepsilon+1)}\vspace{2mm}
\\
c_{pq}^{-\beta\varepsilon}\vspace{2mm}
\\
c_{pq}^{-\beta(\varepsilon+1)}\vspace{2mm}
\end{pmatrix*}
=\\
\epsilon_{p}
\begin{pmatrix}
S_{pq}^{\beta\varepsilon \beta\varepsilon} & 0 & 0 & 0 \vspace{2mm}
\\
0 & S_{pq}^{\beta(\varepsilon+1) \beta(\varepsilon+1)} & 0 & 0 \vspace{2mm}
\\
0  & 0  & S_{pq}^{-\beta\varepsilon -\beta\varepsilon} & 0 \vspace{2mm}
\\
0 & 0 & 0 & S_{pq}^{-\beta(\varepsilon+1) -\beta(\varepsilon+1)} 
\end{pmatrix}
\begin{pmatrix*}[c]
c_{pq}^{\beta\varepsilon}\vspace{2mm}
\\
c_{pq}^{\beta(\varepsilon+1)}\vspace{2mm}
\\
c_{pq}^{-\beta\varepsilon}\vspace{2mm}
\\
c_{pq}^{-\beta(\varepsilon+1)}\vspace{2mm}
\end{pmatrix*}.
\end{multline}
The matrix elements in Eq. (\ref{eq:23_DHFRFOUREXP}) are denoted by,
\begin{align}\label{eq:24_FOCKMATRIXELEM}
f_{pq}^{\beta\varepsilon\beta'\varepsilon'}=
\left\lbrace
\begin{array}{ll}
\vspace{2.5mm} V_{pq}^{\beta\varepsilon\beta'\varepsilon'}-2c^{2}S_{pq}^{\beta\varepsilon\beta'\varepsilon'}\delta_{\beta\beta'}+J_{pq}^{\beta\varepsilon\beta'\varepsilon'}-K_{pq}^{\beta\varepsilon\beta'\varepsilon'} & \hspace{5mm} \beta=\beta'\vee \varepsilon=\varepsilon'
\\
\vspace{2.5mm} -K_{pq}^{\beta\varepsilon\beta'\varepsilon'} & \hspace{5mm} \beta=\beta'\vee \varepsilon \neq \varepsilon'
\\
\vspace{2.5mm} (-1)^{\varepsilon}\hspace{0.5mm} c \hspace{1mm}\leftidx{^{0}}{T}_{pq}^{\beta\varepsilon\beta'\varepsilon'}-K_{pq}^{\beta\varepsilon\beta'\varepsilon'} & \hspace{5mm} \beta \neq \beta'\vee \varepsilon = \varepsilon'
\\
\vspace{2.5mm} c\hspace{1mm}\leftidx{^{-1}}{T}_{pq}^{\beta\varepsilon\beta'\varepsilon'}-K_{pq}^{\beta\varepsilon\beta'\varepsilon'}  & \hspace{5mm} \beta \neq \beta'\vee \varepsilon < \varepsilon'
\\
\vspace{2.5mm} c\hspace{1mm}\leftidx{^{1}}{T}_{pq}^{\beta\varepsilon\beta'\varepsilon'}-K_{pq}^{\beta\varepsilon\beta'\varepsilon'} & \hspace{5mm} \beta \neq \beta'\vee \varepsilon > \varepsilon'
\end{array}
\right.
\end{align}
\twocolumngrid
where, $S_{pq}^{\beta\varepsilon\beta'\varepsilon'}$, $T_{pq}^{\beta\varepsilon\beta'\varepsilon'}$ are overlap and kinetic energy matrices,
\begin{align}\label{eq:25_COULOMBMAT}
J_{pq}^{\beta\varepsilon\beta\varepsilon}
=\sum_{\mu rs}d_{rs}^{\beta\varepsilon_{\mu}\beta\varepsilon_{\mu}}
J_{pqrs}^{\beta\varepsilon\beta\varepsilon\beta\varepsilon_{\mu}\beta\varepsilon_{\mu}},
\end{align}
are two$-$electron Coulomb interaction matrices,
\begin{align}\label{eq:26_EXCHANGEMAT}
K_{pq}^{\beta\varepsilon\beta'\varepsilon'}
=\sum_{rs}d_{rs}^{\beta\varepsilon\beta'\varepsilon'}
K_{pqrs}^{\beta\varepsilon\beta'\varepsilon'\beta\varepsilon\beta'\varepsilon'},
\end{align}
are two$-$electron exchange interaction matrices, and,
\begin{align}\label{eq:27_DENSITYMAT}
d_{rs}^{\beta\varepsilon\beta'\varepsilon'}
=\sum_{i}
c_{ip}^{\beta\varepsilon^{\dagger}}c_{iq}^{\beta'\varepsilon}
\end{align}
are density matrices, $c_{ip}^{\beta\varepsilon^{\dagger}}$ is the complex conjugte of $c_{ip}^{\beta\varepsilon}$, $\mu=\left\lbrace 0,1 \right\rbrace$, $\left\lbrace \varepsilon, \varepsilon' \right\rbrace=\left\lbrace 0, 1 \right\rbrace$.\\
Once the matrix elements given above are evaluated with an initially chosen basis$-$set the methods employed for solution of Eq. (\ref{eq:9_DHFRFOUREXP1}) in non$-$relativistic calculations can readily be adapted to relativistic calculations. The procedures for transformation to an ortho-normal space and computing the eigenvalues such as L{{\"o}}wdin orthogonalization \cite{89_Lowdin_1950}, Cholesky decomposition \cite{90_Press_1992} or Schur decomposition \cite{91_Schur_1909} varies according to the size of matrix, programming language to be used which is also a matter for computer science.

All above matrix elements involve one$-$ and two$-$electron operators up to a maximum three$-$ and four$-$center integrals, respectively. In the Fig. \ref{fig:GeometryFig} depiction of coordinates are given for motion of two$-$electron in a field of four stationary Coulomb centers, where $a$, $b$, $c$, $d$ arbitrary four$-$points of Euclidian space, $\vec{R}_{ab}=\vec{ab}$, $\vec{R}_{ac}=\vec{ac}$, $\vec{r}_{1}=\vec{O1}$, $\vec{r}_{2}=\vec{O2}$, $\vec{r}_{12}=\vec{r}_{1}-\vec{r}_{2}$, $\vec{r}_{a1}=\vec{r}_{1}-\vec{R}_{a}$, $\vec{r}_{a2}=\vec{r}_{2}-\vec{R}_{a}$, and so on. The matrix elements given in Eq. (\ref{eq:24_FOCKMATRIXELEM}) appear in four general forms: overlap integrals $\left(S\right)$, nuclear attraction integrals $\left(V\right)$, kinetic energy integrals $\left(T\right)$, and repulsion integrals, namely Coulomb $\left(J\right)$, exchange $\left(K\right)$ integrals. These integrals can be expressed in terms of non$-$relativistic$-$type molecular integrals as follows \cite{51_Bagci_2016},\\
the overlap and kinetic energy integrals, which are one$-$ or two$-$center integrals,
\begin{multline} \label{eq:28_TCOVERLAP}
S_{nljm,n'l'j'm'}^{\beta\varepsilon \beta'\varepsilon'} \left(\zeta_{a}, \zeta_{b}, \vec{R}_{ab} \right) \\
=\int_{}^{}\chi_{nljm}^{\beta\varepsilon *}\left(\zeta,\vec{r}_{a} \right)\chi_{n'l'j'm'}^{\beta'\varepsilon'}\left(\zeta',\vec{r}_{b} \right)dV\\
=\mathcal{N}^{\beta}_{nj}\left(\zeta_{a} \right)\mathcal{N}^{\beta'}_{n'j'}\left(\zeta_{b} \right)
\text{\large $a$}_{l_{\beta}jm\left( \varepsilon \right)}
\text{\large $\eta$}_{m\left( \varepsilon \right)}
\text{\large $a$}_{l'_{\beta'}jm'\left( \varepsilon' \right)}
\text{\large $\eta$}_{m'\left( \varepsilon' \right)}\\
\times X^{\beta\beta' \dagger}S_{\varepsilon\varepsilon'},
\end{multline}
\begin{multline}\label{eq:29_TCKINETIC}
\leftidx{^{i}}{\hat{T}}_{nljm,n'l'j'm'}^{\beta\varepsilon \beta'\varepsilon'} \left(\zeta_{a}, \zeta_{b}, \vec{R}_{ab} \right) \\
=\int_{}^{}\chi_{nljm}^{\beta\varepsilon *}\left(\zeta,\vec{r}_{a} \right)
\hspace{0.5mm}
\leftidx{^{i}}{\hat{T}}
\hspace{0.5mm}
\chi_{n'l'j'm'}^{\beta'\varepsilon'}\left(\zeta',\vec{r}_{b} \right)dV.
\end{multline}
The kinetic energy integrals can easily be expressed in terms of the overlap and following nuclear attraction integrals i.e., up to a maximum three$-$center integrals, 
\begin{multline} \label{eq:30_TCNUCATTRACT}
\leftidx{^{abc}}{V}_{nljm,n'l'j'm'}^{\beta\varepsilon \beta'\varepsilon'} \left(\zeta_{a}, \zeta_{b}, \vec{R}_{ab}, \vec{R}_{ac} \right) \\
=\int_{}^{}\chi_{nljm}^{\beta\varepsilon *}\left(\zeta,\vec{r}_{a} \right)
\frac{1}{r_{c}}
\chi_{n'l'j'm'}^{\beta'\varepsilon'}\left(\zeta',\vec{r}_{b} \right)dV\\
=\mathcal{N}^{\beta}_{nj}\left(\zeta_{a} \right)\mathcal{N}^{\beta'}_{n'j'}\left(\zeta_{b} \right)
\text{\large $a$}_{l_{\beta}jm\left( \varepsilon \right)}
\text{\large $\eta$}_{m\left( \varepsilon \right)}
\text{\large $a$}_{l'_{\beta'}jm'\left( \varepsilon' \right)}
\text{\large $\eta$}_{m'\left( \varepsilon' \right)}\\
\times X^{\beta\beta' \dagger}
\hspace{1mm}
V_{\varepsilon\varepsilon'},
\end{multline}
where,
\begin{align}\label{eq:31_NROVERLAPMAT}
S_{\varepsilon\varepsilon'}=
\begin{bmatrix}
S_{nlm_{\varepsilon},n'l'm'_{\varepsilon'}}\left(\zeta_{a}, \zeta_{b}, \vec{R}_{ab}\right)
\\
S_{n+1lm_{\varepsilon},n'l'm'_{\varepsilon'}}\left(\zeta_{a}, \zeta_{b}, \vec{R}_{ab}\right)
\\
S_{nlm_{\varepsilon},n'+1l'm'_{\varepsilon'}}\left(\zeta_{a}, \zeta_{b}, \vec{R}_{ab}\right)
\\
S_{n+1lm_{\varepsilon},n'+1l'm'_{\varepsilon'}}\left(\zeta_{a}, \zeta_{b}, \vec{R}_{ab}\right)
\end{bmatrix},
\end{align}
\begin{align}\label{eq:32_NRNUCATTRACTMAT}
V_{\varepsilon\varepsilon'}=
\begin{bmatrix}
V_{nlm_{\varepsilon},n'l'm'_{\varepsilon'}}\left(\zeta_{a}, \zeta_{b}, \vec{R}_{ab}, \vec{R}_{ac} \right)
\\
V_{n+1lm_{\varepsilon},n'l'm'_{\varepsilon'}}\left(\zeta_{a}, \zeta_{b}, \vec{R}_{ab}, \vec{R}_{ac} \right)
\\
V_{nlm_{\varepsilon},n'+1l'm'_{\varepsilon'}}\left(\zeta_{a}, \zeta_{b}, \vec{R}_{ab}, \vec{R}_{ac} \right)
\\
V_{n+1lm_{\varepsilon},n'+1l'm'_{\varepsilon'}}\left(\zeta_{a}, \zeta_{b}, \vec{R}_{ab}, \vec{R}_{ac} \right)
\end{bmatrix},
\end{align}
\begin{align}\label{eq:33_RCOFFMAT}
X^{\beta\beta'}=
\begin{bmatrix}
A^{\beta}_{nlj}A^{\beta'}_{n'l'j'}
\\
\zeta_{B}B^{\beta}_{nlj}A^{\beta'}_{n'l'j'}
\\
\zeta_{B}A^{\beta}_{nlj}B^{\beta'}_{n'l'j'}
\\
\zeta^{2}_{B}B^{\beta}_{nlj}B^{\beta'}_{n'l'j'}
\end{bmatrix},
\end{align} 
are the matrices corresponding to the non$-$relativistic two$-$center overlap and nuclear attraction integrals over Slater$-$type orbitals, coefficients of Slater$-$type spinor orbitals $A^{\beta}_{nlj}$ and $B^{\beta}_{nlj}$, respectively.\\
The Coulomb and exchange matrix elements to be evaluated is, hence of the general form,
\begin{multline}\label{eq:34_FOURCENTERCOULOMB}
J_{pqrs}^{\beta\varepsilon\beta\varepsilon\beta'\varepsilon'\beta'\varepsilon'}\\
=
\iint
\chi_{p}^{\beta\varepsilon*}(\vec{r}_{a1})
\left(
\chi_{r}^{\beta'\varepsilon'*}(\vec{r}_{b2})
\hat{f}_{12}
\chi_{s}^{\beta'\varepsilon'}(\vec{r}_{d2})
\right)
\chi_{p}^{\beta\varepsilon}(\vec{r}_{c1})
dV_{12}\\
=
\mathcal{N}^{\beta}_{n_{a}j_{a}}\left(\zeta_{a} \right)
\mathcal{N}^{\beta}_{n_{c}j_{c}}\left(\zeta_{c} \right)
\mathcal{N}^{\beta'}_{n_{b}j_{b}}\left(\zeta_{b} \right)
\mathcal{N}^{\beta'}_{n_{d}j_{d}}\left(\zeta_{d} \right) \\
\times
\text{\large $\eta$}_{m_{a}\left( \varepsilon \right)}
\text{\large $\eta$}_{m_{b}\left( \varepsilon \right)}
\text{\large $\eta$}_{m'_{c}\left( \varepsilon' \right)}
\text{\large $\eta$}_{m'_{d}\left( \varepsilon' \right)} \\
\times
\text{\large $a$}_{l_{{a}_{\beta}}j_{a}m_{a}\left( \varepsilon \right)}
\text{\large $a$}_{l_{{c}_{\beta}}j_{c}m_{c}\left( \varepsilon \right)}
\text{\large $a$}_{l'_{{b}_{\beta'}}j'_{b}m'_{b}\left( \varepsilon' \right)}
\text{\large $a$}_{l'_{{d}_{\beta'}}j'_{d}m'_{d}\left( \varepsilon' \right)} \\
\times
X_{pq}^{\beta\beta \dagger}
X_{rs}^{\beta'\beta' \dagger}
\hspace{1mm}
J_{\varepsilon\varepsilon'}
\end{multline}
\begin{multline}\label{eq:35_FOURCENTEREXCHANGE}
K_{pqrs}^{\beta\varepsilon\beta'\varepsilon'\beta\varepsilon\beta'\varepsilon'}\\
=
\iint
\chi_{p}^{\beta\varepsilon*}(\vec{r}_{a1})
\left(
\chi_{r}^{\beta\varepsilon*}(\vec{r}_{b2})
\hat{f}_{12}
\chi_{s}^{\beta'\varepsilon'}(\vec{r}_{d1})
\right)
\chi_{p}^{\beta'\varepsilon'}(\vec{r}_{c2})
dV_{12}\\
=
\mathcal{N}^{\beta}_{n_{a}j_{a}}\left(\zeta_{a} \right)
\mathcal{N}^{\beta'}_{n_{c}j_{c}}\left(\zeta_{c} \right)
\mathcal{N}^{\beta}_{n_{b}j_{b}}\left(\zeta_{b} \right)
\mathcal{N}^{\beta'}_{n_{d}j_{d}}\left(\zeta_{d} \right) \\
\times
\text{\large $\eta$}_{m_{a}\left( \varepsilon \right)}
\text{\large $\eta$}_{m_{b}\left( \varepsilon' \right)}
\text{\large $\eta$}_{m'_{c}\left( \varepsilon \right)}
\text{\large $\eta$}_{m'_{d}\left( \varepsilon' \right)} \\
\times
\text{\large $a$}_{l_{{a}_{\beta}}j_{a}m_{a}\left( \varepsilon \right)}
\text{\large $a$}_{l_{{c}_{\beta'}}j_{c}m_{c}\left( \varepsilon' \right)}
\text{\large $a$}_{l'_{{b}_{\beta}}j'_{b}m'_{b}\left( \varepsilon \right)}
\text{\large $a$}_{l'_{{d}_{\beta'}}j'_{d}m'_{d}\left( \varepsilon' \right)} \\
\times
X_{pq}^{\beta\beta' \dagger}
X_{rs}^{\beta\beta' \dagger}
\hspace{1mm}
K_{\varepsilon\varepsilon'}
\end{multline}
 here,
 $p = \left\lbrace n_{a}l_{a}j_{a}m_{a} \right\rbrace$,  $q = \left\lbrace n_{c}l_{c}j_{c}m_{c} \right\rbrace$,  $r = \left\lbrace n_{b}l_{b}j_{b}m_{b} \right\rbrace$,  $s = \left\lbrace n_{d}l_{d}j_{d}m_{d} \right\rbrace$, $dV_{12}=dV_{1}dV_{2}$. The $J_{\varepsilon\varepsilon'}$ and $K_{\varepsilon\varepsilon'}$ matrices are $\left[1 \times 16 \right]$ column matrices whose component are the integrals over non$-$relativistic Slater$-$type orbitals and they are obtained similarly to Eq. (\ref{eq:31_NROVERLAPMAT}) and Eq. (\ref{eq:32_NRNUCATTRACTMAT}).
\section{\label{sec:NonrelatMolInteg}Analytical evaluation for non$-$relativistic molecular integrals}
\begin{table*}
\caption{\label{tab:HlikeECT} Results of computation for some electronic energy states of H$-$like ions depending on nuclear charge $Z$.}
\begin{ruledtabular}
\begin{tabular}{ccccccc}
Radial exponent&$Z$ &$1s_{1/2}$ &$2s_{1/2}$ &$2p_{1/2}$ &$2p_{3/2}$ &$3s_{1/2}$
\\
$n=\vert \kappa \vert +0.0$
&
\begin{tabular}[c]{@{}l@{}l@{}l@{}l@{}l@{}l@{}l@{}}
110 \\
120 \\
130 \\
136 \\
137 \\
138 \\
139 \\
140 \\
150 \\
160
\end{tabular}
& \begin{tabular}[c]{@{}l@{}l@{}l@{}l@{}l@{}l@{}l@{}}
\begin{tabular}{cc}
204.90 & 204.87
\end{tabular} \\
\begin{tabular}{cc}
260.00 & 259.60
\end{tabular} \\
\begin{tabular}{cc}
332.66 & 330.43
\end{tabular} \\
\begin{tabular}{cc}
390.01 & {}
\end{tabular} \\
\begin{tabular}{cc}
401.04 & {}
\end{tabular} \\
\begin{tabular}{cc}
412.60 & {}
\end{tabular} \\
\begin{tabular}{cc}
424.74 & {}
\end{tabular} \\
\begin{tabular}{cc}
437.52 & 425.89
\end{tabular} \\
\\
{}
\end{tabular}
& \begin{tabular}[c]{@{}l@{}l@{}l@{}l@{}l@{}l@{}l@{}}
\begin{tabular}{cc}
54.561 & 54.150
\end{tabular} \\
\begin{tabular}{cc}
70.629 & 69.780
\end{tabular} \\
\begin{tabular}{cc}
93.266 & 90.790
\end{tabular} \\
\begin{tabular}{cc}
112.78 & {}
\end{tabular} \\
\begin{tabular}{cc}
116.75 & {}
\end{tabular} \\
\begin{tabular}{cc}
121.01 & {}
\end{tabular} \\
\begin{tabular}{cc}
125.60 & {}
\end{tabular} \\
\begin{tabular}{cc}
130.55 & 120.11
\end{tabular} \\
\\
{}
\end{tabular} 
& \begin{tabular}[c]{@{}l@{}l@{}l@{}l@{}l@{}l@{}l@{}}
\begin{tabular}{cc}
54.228 & 54.400
\end{tabular} \\
\begin{tabular}{cc}
70.160 & 70.640
\end{tabular} \\
\begin{tabular}{cc}
92.569 & 93.840
\end{tabular} \\
\begin{tabular}{cc}
111.81 & {}
\end{tabular} \\
\begin{tabular}{cc}
115.72 & {}
\end{tabular} \\
\begin{tabular}{cc}
119.89 & {}
\end{tabular} \\
\begin{tabular}{cc}
124.37 & {}
\end{tabular} \\
\begin{tabular}{cc}
129.18 & 131.4
\end{tabular} \\
\\
{}
\end{tabular} 
& \begin{tabular}[c]{@{}l@{}l@{}l@{}l@{}l@{}l@{}l@{}}
\begin{tabular}{cc}
42.963 & 42.970
\end{tabular} \\
\begin{tabular}{cc}
51.584 & 51.590
\end{tabular} \\
\begin{tabular}{cc}
61.142 & 69.150
\end{tabular} \\
\begin{tabular}{cc}
67.351 & {}
\end{tabular} \\
\begin{tabular}{cc}
68.422 & {}
\end{tabular} \\
\begin{tabular}{cc}
69.503 & {}
\end{tabular} \\
\begin{tabular}{cc}
70.595 & {}
\end{tabular} \\
\begin{tabular}{cc}
71.698 & 71.700
\end{tabular} \\
\begin{tabular}{cc}
83.236 & 83.330
\end{tabular} \\
\begin{tabular}{cc}
96.116 & 96.120
\end{tabular} 
\end{tabular} 
& \begin{tabular}[c]{@{}l@{}l@{}l@{}l@{}l@{}l@{}l@{}}
\begin{tabular}{cc}
22.679 & 23.700
\end{tabular} \\
\begin{tabular}{cc}
28.520 & 28.740
\end{tabular} \\
\begin{tabular}{cc}
35.817 & 36.490
\end{tabular} \\
\begin{tabular}{cc}
41.126 & {}
\end{tabular} \\
\begin{tabular}{cc}
42.093 & {}
\end{tabular} \\
\begin{tabular}{cc}
43.086 & {}
\end{tabular} \\
\begin{tabular}{cc}
44.105 & {}
\end{tabular} \\
\begin{tabular}{cc}
45.153 & 46.760
\end{tabular} \\
\begin{tabular}{cc}
57.405 & 60.450
\end{tabular} \\
\begin{tabular}{cc}
73.814  & 78.090
\end{tabular}
\end{tabular} \\
\hline
$n=\vert \kappa \vert +0.3$ 
&
\begin{tabular}[c]{@{}l@{}l@{}l@{}l@{}l@{}l@{}l@{}}
150 \\
160 
\end{tabular}
& \begin{tabular}[c]{@{}l@{}l@{}l@{}l@{}l@{}l@{}l@{}}
\begin{tabular}{cc}
559.84 & 559.76
\end{tabular} \\
{}
\end{tabular}
& \begin{tabular}[c]{@{}l@{}l@{}l@{}l@{}l@{}l@{}l@{}}
\begin{tabular}{cc}
192.97 & 161.58
\end{tabular} \\
{}
\end{tabular}
& \begin{tabular}[c]{@{}l@{}l@{}l@{}l@{}l@{}l@{}l@{}}
\begin{tabular}{cc}
187.56 & 201.63
\end{tabular} \\
{}
\end{tabular} 
& \begin{tabular}[c]{@{}l@{}l@{}l@{}l@{}l@{}l@{}l@{}}
\begin{tabular}{cc}
83.236 & 83.330
\end{tabular} \\
\begin{tabular}{cc}
96.116 & 96.120
\end{tabular}
\end{tabular} 
& \begin{tabular}[c]{@{}l@{}l@{}l@{}l@{}l@{}l@{}l@{}}
\begin{tabular}{cc}
52.800 & 60.450
\end{tabular} \\
\begin{tabular}{cc}
65.714 & 78.090
\end{tabular}
\end{tabular} \\
\hline
$n=\vert \kappa \vert + 0.6$
& 
\begin{tabular}[c]{@{}l@{}l@{}l@{}l@{}l@{}l@{}l@{}}
150 \\
160 
\end{tabular}
& \begin{tabular}[c]{@{}l@{}l@{}l@{}l@{}l@{}l@{}l@{}}
\begin{tabular}{cc}
514.25 & 559.76
\end{tabular} \\
{}
\end{tabular} 
& \begin{tabular}[c]{@{}l@{}l@{}l@{}l@{}l@{}l@{}l@{}}
\begin{tabular}{cc}
170.31 & 161.58
\end{tabular} \\
{}
\end{tabular} 
& \begin{tabular}[c]{@{}l@{}l@{}l@{}l@{}l@{}l@{}l@{}}
\begin{tabular}{cc}
167.65 & 201.63
\end{tabular} \\
{}
\end{tabular} 
& \begin{tabular}[c]{@{}l@{}l@{}l@{}l@{}l@{}l@{}l@{}}
\begin{tabular}{cc}
83.236 & 83.330
\end{tabular} \\
\begin{tabular}{cc}
96.116 & 96.120
\end{tabular}
\end{tabular} 
& \begin{tabular}[c]{@{}l@{}l@{}l@{}l@{}l@{}l@{}l@{}}
\begin{tabular}{cc}
49.364 & 60.450
\end{tabular} \\
\begin{tabular}{cc}
60.256 & 78.090
\end{tabular}
\end{tabular} \\
\hline
$n=\vert \kappa \vert + 0.9$ 
&
\begin{tabular}[c]{@{}l@{}l@{}l@{}l@{}l@{}l@{}l@{}}
150 \\
160 
\end{tabular}
& \begin{tabular}[c]{@{}l@{}l@{}l@{}l@{}l@{}l@{}l@{}}
\begin{tabular}{cc}
481.45 & 559.76
\end{tabular} \\
{}
\end{tabular} 
& \begin{tabular}[c]{@{}l@{}l@{}l@{}l@{}l@{}l@{}l@{}}
\begin{tabular}{cc}
156.67 & 161.58
\end{tabular} \\
{}  
\end{tabular} 
& \begin{tabular}[c]{@{}l@{}l@{}l@{}l@{}l@{}l@{}l@{}}
\begin{tabular}{cc}
154.93 & 201.63
\end{tabular} \\
{}
\end{tabular} 
& \begin{tabular}[c]{@{}l@{}l@{}l@{}l@{}l@{}l@{}l@{}}
\begin{tabular}{cc}
83.236 & 83.330
\end{tabular} \\
\begin{tabular}{cc}
96.116 & 96.120
\end{tabular} 
\end{tabular} 
& \begin{tabular}[c]{@{}l@{}l@{}l@{}l@{}l@{}l@{}l@{}}
\begin{tabular}{cc}
46.594 & 60.450
\end{tabular} \\
\begin{tabular}{cc}
56.152 & 78.090
\end{tabular}
\end{tabular} \\
\hline
$n=2\vert \kappa \vert$ 
&
\begin{tabular}[c]{@{}l@{}l@{}l@{}l@{}l@{}l@{}l@{}}
150 \\
160 
\end{tabular}
& \begin{tabular}[c]{@{}l@{}l@{}l@{}l@{}l@{}l@{}l@{}}
\begin{tabular}{cc}
439.32 & 559.76
\end{tabular} \\
\begin{tabular}{cc}
560.32 & 749.68
\end{tabular} 
\end{tabular} 
& \begin{tabular}[c]{@{}l@{}l@{}l@{}l@{}l@{}l@{}l@{}}
\begin{tabular}{cc}
141.56 & 161.58
\end{tabular} \\
\begin{tabular}{cc}
202.74 & 217.54
\end{tabular}
\end{tabular}
& \begin{tabular}[c]{@{}l@{}l@{}l@{}l@{}l@{}l@{}l@{}}
\begin{tabular}{cc}
140.35 & 201.63
\end{tabular} \\
\begin{tabular}{cc}
202.37 & 336.25
\end{tabular}  
\end{tabular}
& \begin{tabular}[c]{@{}l@{}l@{}l@{}l@{}l@{}l@{}l@{}}
\begin{tabular}{cc}
83.236 & 83.330
\end{tabular} \\
\begin{tabular}{cc}
96.116 & 96.120
\end{tabular}
\end{tabular}
& \begin{tabular}[c]{@{}l@{}l@{}l@{}l@{}l@{}l@{}l@{}}
\begin{tabular}{cc}
43.810 & 60.450
\end{tabular} \\
\begin{tabular}{cc}
52.126 & 78.090
\end{tabular}
\end{tabular}
\end{tabular}
\footnotetext{The values on second row indicates results obtained via numerical solution \cite{113_Pieper_1969}. There the nucleus radii were determined from the relationship $r_{N}=1.2X10^{-3}A^{1/3}cm$}
\footnotetext{Absolute value of the electronic energy states are given in KeV.}
\end{ruledtabular}
\end{table*}
The corresponding non$-$relativistic matrix elements of $\left(S_{\varepsilon\varepsilon'}, V_{\varepsilon\varepsilon'}, J_{\varepsilon\varepsilon'}, K_{\varepsilon\varepsilon'} \right)$ through Laplace expansion of Coulomb interaction and prolate spheroidal coordinates explicitly are given in lined$-$up coordinate system by the following formulas \citep{51_Bagci_2016,67_Bagci_2014,68_Bagci_2015,69_Bagci_2015},\\
for two$-$center overlap,
\begin{multline}\label{eq:36_NROVERLAPEXPLOCAL}
S_{nl\lambda,n'l'\lambda}\left(\zeta_{a}, \zeta_{b}, R_{ab}\right)
=\sum_{\alpha=0}^{l}\sum_{\beta=\lambda}^{l'}\sum_{q=0}^{a+b}g_{\alpha\beta}^{q}\left(l\lambda,l'\lambda \right)\\
\times \mathcal{P}^{0,q}_{n-a\alpha,n'-\beta,0}\left(0,\frac{R_{ab}}{2}\left(\zeta_{a}+\zeta_{b} \right),\frac{R_{ab}}{2}\left(\zeta_{a}-\zeta_{b} \right) \right),
\end{multline}
and nuclear attraction integrals,
\begin{multline}\label{eq:37_NRNUCATTRACTEXP}
V_{nl\lambda,n'l'\lambda}\left(\zeta_{a},\zeta'_{a}, R_{ab}\right)\\
=\sum_{L}\sqrt{\frac{4\pi}{2L+1}}C^{L0}(l\lambda,l'\lambda)\\
\times R^{L}_{nn'}\left(\zeta_{a},\zeta'_{a}, R_{ab} \right)Y^{*}_{L0}\left(0, 0 \right),
\end{multline}
where, $\lambda=\vert m_{l} \vert =\vert m'_{l} \vert$, $R_{n,n'}^{L}$ is the single-center potential,
\begin{multline}\label{eq:38_RADIALPOTFUNCS}
	R_{n,n'}^{L} \left(\zeta_{a},\zeta'_{a}, R_{ab} \right)\\
	= \left(2\overline{\zeta_{a}} \right)\Gamma\left (n+n'+L+1 \right)\frac{1}{\left(2\overline{\zeta_{a}}R_{ab} \right)^{L+1}} \\
	\times\Bigg\{P \left [n+n'+L+1,2\overline{\zeta_{a}}R_{ab} \right] \Bigg. \\+\left. \frac{\left(2\overline{\zeta_{a}}R_{ab} \right)^{2L+1}}{\left(n+n'-L\right)_{2L+1}}Q \left [n+n'-L,2\overline{\zeta_{a}}R_{ab} \right] \right\},
\end{multline}
$g_{\alpha\beta}^{q}$ coefficients arise from product of two spherical harmonics with different centers \cite{92_Guseinov_1970},
\begin{align} \label{eq:39_GABC1}
	g_{\alpha\beta}^{q}(l\lambda,l'\lambda)=g_{\alpha\beta}^{0}(l\lambda,l'\lambda)F_{q}(\alpha+\lambda,\beta-\lambda)
\end{align}
\begin{align} \label{eq:40_GABC2}
	g_{\alpha\beta}^{0}(l\lambda,l'\lambda)=\sum_{s=0}^{\nu}(-1)^{s}F_{s}(\lambda)D_{\alpha+2\lambda-2s}^{l\lambda}D_{\beta}^{l'\lambda},
\end{align}
\begin{multline} \label{eq:41_DBLNU}
	D_{b}^{l\lambda}=\frac{1}{2^l}(-1)^{(l-b)/2}\left[\frac{2l+1}{2}\frac{F_{l}(l+\lambda)}{F_{\lambda}(l)} \right]^{1/2}\\
	\times F_{(l-\beta)/2}(l)F_{\beta-\lambda}(l+\beta),
\end{multline}
with, the quantities $F_{s}(n,n')$ are the generalized binomial coefficients and they are given as,
\begin{align} \label{eq:42_GENBINOM}
	F_{s}(n,n')=\sum_{s'}(-1)^{s'}F_{s-s'}(n)F_{s'}(n')
\end{align}
and, $\left\lbrace n, n' \right\rbrace \in \mathbb{Z}^{+}$, $\frac{1}{2}\left[(s-n)+\vert s-n \vert \right]\leq s' \leq min(s,n)$\\
$Q\left[\alpha,x \right]$, $\Gamma\left[\alpha,x \right]$ is the normalized complementary incomplete gamma, complementary incomplete gamma functions,
\begin{align}\label{eq:43_COMPNORMINCOMPGAMMA}
	Q\left[\alpha,x \right]=\frac{\Gamma(\alpha, x)}{\Gamma(\alpha)},
\end{align}
\begin{align}\label{eq:44_COMPINCOMPGAMMA}
\Gamma\left(\alpha,x \right)=
\int_{x}^{\infty}t^{\alpha-1}e^{-t}dt.
\end{align}
Due to wide range of use in applied science accurate calculation of incomplete gamma functions is one of the most important topic in modern analysis. \cite{93_Chaudhry_2002, 94_Gautschi_2003, 95_Blahak_2010}. An efficient approach for computing the incomplete gamma functions without erroneous last digits is still being studied in the literature \cite{93_Chaudhry_2002, 96_Gautschi_1979, 97_Temme_1994_1, 98_Temme_1994_2}. Several methods are available. Four domains of computation for the incomplete gamma functions ratios corresponding to these methods were indicated in \cite{96_Gautschi_1979, 99_Gil_2012}. The domains were established as a compromise between efficiency and accuracy.\\
Convergence behavior of the incomplete gamma functions may be predicted by a method given in \cite{99_Gil_2012}. To estimate the number of terms that are needed to achieve a certain accuracy after truncating the series,\\
it is written,
\begin{equation}\label{eq:45_ALTERGAMMA1}
\sum_{s=0}^{\infty}\frac{x^{s}}{\left(a+1\right)_{s}}
=S_{s_{0}}\left(a, x\right)+R_{s_{0}}\left(a,x \right),
\end{equation}
where,
\begin{equation}\label{eq:46_ALTERGAMMA1}
S_{s_{0}}\left(a, x\right)=\sum_{s=0}^{s_{0}-1}\frac{x^{s}}{\left(a+1\right)_{s}},
\hspace{5mm} R_{s_{0}}\left(a, x\right)=\sum_{s=s_{0}}^{\infty}\frac{x^{s}}{\left(a+1\right)_{s}},
\end{equation}
and it is computed the smallest $s=s_{0}$ that satisfies,
\begin{equation}\label{eq:47_ALTERGAMMA1}
\frac{x^{s}}{\left(a+1\right)_{s}}\leq \epsilon.
\end{equation}

Compact expressions for the two$-$center two$-$electron Coulomb and hybrid integrals are obtained by generalizing the solution of the Poisson equation as a partial differential equation in spherical coordinates by expanding the potential the set of functions referred to as spectral forms (SFs) \cite{100_Weatherford_2005, 101_Absi_2006}. Through Laplace expansion of Coulomb interaction and prolate$-$spheroidal coordinates the radial parts of these integrals are expressed in terms of upper $\left(\mathcal{P}\right)-$ and lower $\left(\mathcal{Q}\right)-$components of relativistic molecular auxiliary functions as follows \cite{68_Bagci_2015},\\
The two$-$center Coulomb integrals,
\begin{widetext}
\begin{multline} \label{eq:48_TCI2}
\mathcal{J}^{aa,bb}_{n_{1}l_{1}m_{1},n_{1}'l_{1}'m_{1}';n_{1}l_{2}m_{2},n_{2}'l_{2}'m_{2}'}
\left (\zeta_{1},\zeta_{1}';\zeta_{2},\zeta_{2}' \right) \\
=\frac{2}{R}\mathcal{N}_{n_{1}n_{1}'}(1,t_{1})\mathcal{N}_{n_{2}n_{2}'}(p_{2},t_{2})\times
\sum_{L_{1}L_{2}M}{}\left(\frac{2L_{2}+1}{2L_{1}+1}\right)A_{m_{1}m_{1}'}^{M}A_{m_{2}m_{2}'}^{M}C^{L_{1}M}(l_{1}m_{1};l_{1}'m_{1}')C^{L_{2}M}(l_{2}m_{2};l_{2}'m_{2}') \\
\times \Gamma(n_{1}+n_{1}'+L_{1}+1)\frac{1}{p_{1}^{L_{1}}} \sum_{\alpha\beta q}g_{\alpha\beta}^{q}(L_{1}\lambda,L_{2}\lambda)\\
\times \left\{ \mathcal{P}_{\sl L_{1}+\alpha, \sl n_{2}+n_{2}'-\beta-1, \sl n_{1}+n_{1}'+L_{1}+1}^{0,q} \left(p_{1},p_{2},-p_{2} \right)+ \mathcal{Q}_{\sl \alpha-(L_{1}+1), \sl n_{2}+n_{2}'-\beta-1, \sl n_{1}+n_{1}'+L_{1}+1}^{2L_{1}+1,q} \left(p_{1},p_{2},-p_{2} \right) \right\},
\end{multline}
$\mathit{max}[\left|{-L_{1},-L_{2}}\right|]\le M\le
\mathit{min}[L_{1}+L_{2}],$ $\left|{l_{1}-l_{1}'}\right|\le L_{1}\le
l_{1}+l_{1}',$ $\left|{l_{2}-l_{2}'}\right|\le L_{2}\le l_{2}+l_{2}'$.
And, the two$-$center hybrid integrals,
\begin{multline} \label{eq:49_THI2}
\mathcal{H}^{aa,ab}_{n_{1}l_{1}m_{1},n_{1}'l_{1}'m_{1}';n_{1}l_{2}m_{2},n_{2}'l_{2}'m_{2}'}
\left (\zeta_{1},\zeta_{1}';\zeta_{2},\zeta_{2}' \right) \\
=\frac{2}{R}\mathcal{N}_{n_{1}n_{1}'}(1,t_{1})
\mathcal{N}_{n_{2}n_{2}'}(p_{2},t_{2})\times\sum_{L_{1}M_{1}L_{2}}{}\left(\frac{2L_{2}+1}{2L_{1}+1}\right)
A_{m_{1}m_{1}'}^{M}
A_{M_{1}m_{2}'}^{m_{2}'} C^{L_{1}M}(l_{1}m_{1};l_{1}'m_{1}')C^{L_{2}m_{2}'}(L_{1}M_{1};l_{2}m_{2}) \\
\times \Gamma(n_{1}+n_{1}'+L_{1}+1)\frac{1}{p_{1}^{L_{1}}} \sum_{\alpha\beta q}g_{\alpha\beta}^{q}(L_{1}\lambda,L_{2}\lambda)\\
\times \left\{ \mathcal{P}_{\sl L_{1}+\alpha +1-n_{2}, \sl n_{2}'-\beta, \sl n_{1}+n_{1}'+L_{1}+1}^{0,q} \left(p_{1},p_{2},p_{2}t_{2} \right)+ \mathcal{Q}_{\sl \alpha-L_{1}-n_{2}, \sl n_{2}'-\beta -1, \sl n_{1}+n_{1}'+L_{1}+1}^{2L_{1}+1,q} \left(p_{1},p_{2},p_{2}t_{2} \right) \right\},
\end{multline}
$\left|{l_{1}-l_{1}'}\right|\le L_{1}\le l_{1}+l_{1}'$
$-L_{1}\le M_{1}\le L_{1},$ $\left|{L_{1}-l_{2}}\right|\le L_{2}\le
L_{1}+l_{2}$.

The auxiliary functions occurring in analytically closed form
expressions given in Eqs.(\ref{eq:48_TCI2}, \ref{eq:49_THI2}) given as,
\begin{multline} \label{eq:50_AUX}
\left\lbrace \begin{array}{cc}
\mathcal{P}^{n_1,q}_{n_{2}n_{3}n_{4}}\left(p_{123} \right)
\\
\mathcal{Q}^{n_1,q}_{n_{2}n_{3}n_{4}}\left(p_{123} \right)
\end{array} \right\rbrace
=\frac{p_{1}^{\sl n_{1}}}{\left({\sl n_{4}}-{\sl n_{1}} \right)_{\sl n_{1}}}
\int_{1}^{\infty}\int_{-1}^{1}{\left(\xi\nu \right)^{q}\left(\xi+\nu \right)^{\sl n_{2}}\left(\xi-\nu \right)^{\sl n_{3}}}
\\
\times \left\lbrace \begin{array}{cc}
P\left[{\sl n_{4}-n_{1}},p_{1}f^{k}_{ij}(\mu,\nu) \right]
\\
Q\left[{\sl n_{4}-n_{1}},p_{1}f^{k}_{ij}(\mu,\nu) \right]
\end{array} \right\rbrace
e^{p_{2}\xi-p_{3}\nu}d\xi d\nu
\end{multline}
with $f^{0}_{10}(\xi,\nu)=\xi+\nu$, $(a)_{n}$ is the Pochhammer symbol, $\left\lbrace q, n_{1} \right\rbrace \in \mathbb{Z}_{0}^{+}$, $\left\lbrace n_{2}, n_{3}, n_{4}\right\rbrace \in \mathbb{R}$, $p_{123}=\left\lbrace p_{1}, p_{2}, p_{3}\right\rbrace$ (and in subsequent notations).
\end{widetext}
The functions $P\left[a, x\right]$, 
\begin{align}\label{eq:51_NORMINCOMPGAMMA}
P\left[\alpha,x \right]=\frac{\gamma(\alpha, x)}{\Gamma(\alpha)},
\end{align}
\begin{align}\label{eq:52_INCOMPGAMMA}
\gamma\left(\alpha,x \right)=
\int_{0}^{x}t^{\alpha-1}e^{-t}dt.
\end{align}
yet to be defined are the normalized incomplete gamma functions and incomplete gamma functions, respectively. Note that, $P$ and $Q$ satisfy the identity $P+Q=1$.

Free Boost {\sl C++} special functions and multi-precision libraries \cite{102_BoostCpp}, together for instance can be used alternatively to Mathematica programming language in order to calculate these functions with high numerical accuracy. Another and more favorable method is to use {\sl Julia} \cite{103_Bezanson_2017} programming language. {\sl Julia} programming language allow easy use of this existing code written in {C} or {\sl Fortran} programming languages. This programming language has a "no boilerplate" philosophy: functions can be called directly from it without any "glue" code, code generation, or compilation even from the interactive prompt. This is accomplished by making an appropriate call with {\sl ccall}, which looks like an ordinary function call.\\
The most common syntax for {\sl ccall} is as follow,
\begin{multline*}
ccall(
(symbol,library), \\
RetType,
(ArgType1, ...), 
Arg1, ... 
).
\end{multline*}
For accuracy only an additional computer algebra package so called {\sl Nemo} \cite{104_Fieker_2017} is required. This package is based on $C$ libraries such as $FLINT, ANTIC, Arb, Pari$ and $Singular$. It has a module system which is use to provide access to $Nemo$. It is imported and used all exported functionality by simply type $using$ $Nemo$.
\section{\label{sec:molauxrev}Evaluation of Relativistic Molecular Auxiliary Integrals}
\begin{figure*}[!]
\centering
\includegraphics[width=1.0\textwidth,height=0.50\textheight]{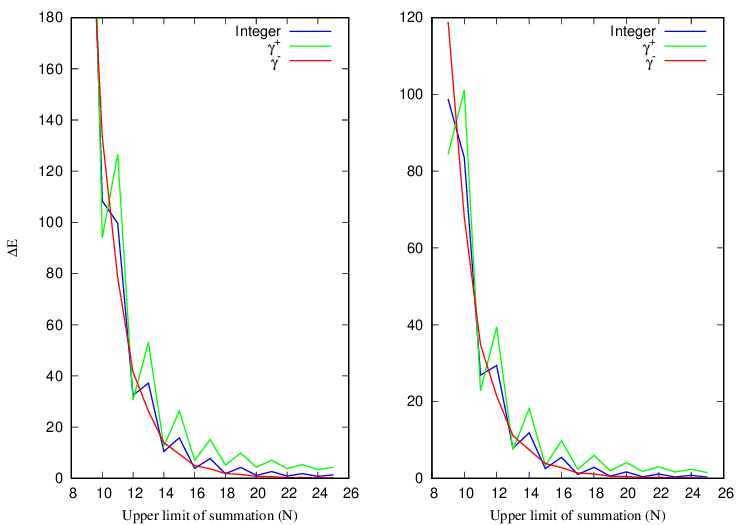}
\caption{Difference between energy eigenvalues $\Delta E=E_{2s_{1/2}}-E_{2p_{1/2}}$ (left), $\Delta E=E_{3s_{1/2}}-E_{3p_{1/2}}$ (right) of the Dirac equation solution, via LCSO method, where the principal quantum numbers are taken to be $n=\gamma^{-}=\sqrt{\kappa^2-{Z^2}/{c^2}}$ (red line), $n \in \mathbb{Z}^{+}\left(n=\vert \kappa \vert \right)$ (blue line), $n=\gamma^{+}=\sqrt{\kappa^2+{Z^2}/{c^2}}$ (green line) and orbital parameters $\zeta=Z$ (for left), $\zeta=Z/2$ (for right) for hydrogen$-$like atom with nuclear charge $Z=50$ in atomic units (a.u.). The results are multiplied by $10^{3}$.}
\label{fig:DIRACSTOEIGEN}
\end{figure*}
Molecular auxiliary functions given in Eq. (\ref{eq:50_AUX}) are among the most challenging integrals in the literature since they involve power functions with non$-$integer exponents, incomplete gamma functions and their products have no explicit closed$-$form relations. The incomplete gamma functions in Eq. (\ref{eq:50_AUX}) arise as a result of two$-$electron interactions. The general form of $f^{k}_{ij}(\xi,\nu)$ represents the interaction potentials which can be generalized to whole set of physical potentials operators as follows:
\begin{align}\label{eq:53_POTOPT}
f^{k}_{ij}(\xi,\nu)
=\left(\xi \nu \right)^{k} \left(\xi+\nu \right)^{i} \left(\xi-\nu\right)^{j}.
\end{align}
The elements in $f^{k}_{ij}$ are irreducible representations required to generate the potential and include the Coulomb potential as a special case when $i=1$, $j=k=0 \hspace{2mm} (f^{0}_{10}(\xi,\nu)=\xi+\nu)$, where $\left\{i,j,k\right\}\in \mathbb{Z}_{0}^{+}$ \cite{68_Bagci_2015}.

The sum of $\mathcal{P}^{n_1,q}$, $\mathcal{Q}^{n_1,q}$ auxiliary functions in Eq. (\ref{eq:50_AUX}) becomes independent from electron$-$electron interactions and reduces to well known auxiliary functions that represent the electron$-$nucleus interaction \cite{92_Guseinov_1970, 105_Pople_1970},
\begin{multline} \label{eq:54_AUXG}
\mathcal{G}^{n_1,q}_{n_{2}n_{3}n_{4}}\left(p_{123} \right)
=\frac{p_{1}^{\sl n_{1}}}{\left({\sl n_{4}}-{\sl n_{1}} \right)_{\sl n_{1}}}
\\
\int_{1}^{\infty}\int_{-1}^{1}{\left(\xi\nu \right)^{q}\left(\xi+\nu \right)^{\sl n_{2}}\left(\xi-\nu \right)^{\sl n_{3}}}
e^{p_{2}\xi-p_{3}\nu}d\xi d\nu.
\end{multline}
This property is quite important since forms of $\mathcal{P}^{n_1,q}$, $\mathcal{Q}^{n_1,q}$ arising in the Eq. (\ref{eq:48_TCI2}) and Eq. (\ref{eq:49_THI2}) are available to reduce to $\mathcal{G}^{n_1,q}$ given in Eq. (\ref{eq:54_AUXG}). Hence, avoiding direct calculation of $\mathcal{P}^{n_1,q}$, $\mathcal{Q}^{n_1,q}$ (and the incomplete gamma functions, consequently).\\
Considering together Eqs. (\ref{eq:48_TCI2}, \ref{eq:49_THI2}) with Eq. (\ref{eq:50_AUX}) and a simple change in Eq. (\ref{eq:50_AUX}) expressing the variable as:
\begin{align*}
\left\lbrace \begin{array}{cc}
\mathcal{P}^{N_1,q}_{N_{2}N_{3}N_{4}}\left(p_{123} \right)
\\
\mathcal{Q}^{N_1,q}_{N_{2}N_{3}N_{4}}\left(p_{123} \right)
\end{array} \right\rbrace
\equiv
\left\lbrace \begin{array}{cc}
\mathcal{P}^{n_1,q}_{n_{2}n_{3}n_{4}}\left(p_{123} \right)
\\
\mathcal{Q}^{n_1,q}_{n_{2}n_{3}n_{4}}\left(p_{123} \right)
\end{array} \right\rbrace,
\end{align*}
it is easy to see that,
\begin{align*}
\left. \begin{array}{cc}
N_{1}=0, \hspace{2mm} N_{4}=n_{1}+n_{1}'+L_{1}+1 \hspace{15mm} for \hspace{5mm}  \mathcal{P}
\\
N_{1}=2L_{1}+1, \hspace{2mm} N_{4}=n_{1}+n_{1}'+L_{1}+1 \hspace{5mm} for \hspace{5mm}  \mathcal{Q}
\end{array} \right.
\end{align*}
and,
\begin{align*}
\left. \begin{array}{cc}
N_{4}-N_{1}=n_{1}+n_{1}'+L_{1}+1 \hspace{5mm} for \hspace{5mm}  \mathcal{P}
\\
N_{4}-N_{1}=n_{1}+n_{1}'-L_{1} \hspace{11mm} for \hspace{5mm}  \mathcal{Q}
\end{array} \right.
\end{align*}
In order to take advantage of sum $\left(P+Q=1 \right)$, $N_{4}-N_{1}$ for both $\mathcal{P}^{N_1,q}$, $\mathcal{Q}^{N_1,q}$ should have same value. Since total angular momentum quantum numbers $L_{1}$ are in set of positive integer numbers $\left( L_{1} \in \mathbb{Z}_{0}^{+} \right)$, it is possible to synchronize $N_{4}-N_{1}$ to the value 
$N_{4}-N_{1}=n_{1}+n_{1}'+L_{1}$
or to that where
$N_{4}-N_{1}=n_{1}+n_{1}'+L_{1}+1$ 
by the following upward and downward distant recurrence relations of $\mathcal{P}^{N_1,q}$ and $\mathcal{Q}^{N_1,q}$,
\begin{multline}\label{eq:55_NOMINCOMPGAMDOWNDREC}
\left\lbrace \begin{array}{cc}
P\left[a,bz \right]
\\
Q\left[a,bz \right]
\end{array} \right\rbrace
\\
=
\left\lbrace \begin{array}{cc}
P\left[a+n,bz \right]+e^{-bz}\sum_{s=1}^{n}\frac{\left(bz\right)^{a+s-1}}{\Gamma(a+s)}
\\
Q\left[a+n,bz \right]-e^{-bz}\sum_{s=1}^{n}\frac{\left(bz\right)^{a+s-1}}{\Gamma(a+s)}
\end{array} \right\rbrace,
\end{multline}
\begin{multline}\label{eq:55_NOMINCOMPGAMDOWNDREC}
\left\lbrace \begin{array}{cc}
P\left[a,bz \right]
\\
Q\left[a,bz \right]
\end{array} \right\rbrace
\\
=
\left\lbrace \begin{array}{cc}
P\left[a-n,bz \right]-e^{-bz}\sum_{s=0}^{n-1}\frac{\left(bz\right)^{a-s-1}}{\Gamma(a+s)}
\\
Q\left[a-n,bz \right]+e^{-bz}\sum_{s=0}^{n-1}\frac{\left(bz\right)^{a-s-1}}{\Gamma(a-s)}
\end{array} \right\rbrace.
\end{multline}
The $\mathcal{G}^{n_1,q}$ auxiliary functions are in fac the representation of two$-$center overlap integrals in prolate$-$spheroidal coordinates. Instead of using the ill$-$conditioned series representation \cite{106_Guseinov_2002} for analytically evaluation them, here  series representation of incomplete beta functions are used,\\
if the parameter $p_{3}=0$;\\
Starting by lowering the indices $q$ for ${\mathcal{G}^{n_{1},q}}$ auxiliary functions,
\begin{equation}\label{eq:57_LOWERINDICES}
\left(\xi\nu \right)=\frac{1}{4}\left\lbrace \left(\xi+\nu \right)^{2}-\left(\xi-\nu \right)^{2} \right\rbrace,
\end{equation}
we have,
\begin{align}\label{eq:58_GRECRELAT}
{\mathcal{G}^{n_{1},q}_{n_{2}n_{3}}}(p_{120})
=\frac{1}{4}
\left\lbrace
{\mathcal{G}^{n_{1},q-1}_{n_{2}+2n_{3}}}(p_{120})
-
{\mathcal{G}^{n_{1},q-1}_{n_{2}n_{3}+2}}(p_{120})
\right\rbrace,
\end{align}
for $q=0$ the expression become,
\begin{multline}\label{eq:59_GALTER1}
\mathcal{G}^{n_{1},0}_{n_{2}n_{3}}\left(p_{120} \right)
=h^{n_{1},0}_{n_{2}n_{3}}\left(p_{12} \right)+ h^{n_{1},0}_{n_{3}n_{2}}\left(p_{12}\right) 
\\
-k^{n_{1},0}_{n_{2}n_{3}}\left(p_{12}\right)
-k^{n_{1},0}_{n_{3}n_{2}}\left(p_{12}\right),
\end{multline}
here,
\begin{multline}\label{eq:60_GALTERH}
h^{n_{1},q}_{n_{2}n_{3}}\left(p_{12}\right)
=\frac{p_{1}^{n_{1}}}{\Gamma\left(n_{1}+1 \right)}2^{n_{2}+n_{3}+1}B\left(n_{2}+1,n_{3}+1\right)
\\
E_{-\left(n_{2}+n_{3}+q+1\right)}\left(p_{2}\right)
-l^{n_{1},q}_{n_{2}n_{3}}\left(p_{12}\right),
\end{multline}
\begin{multline}\label{eq:61_GALTERL}
l^{n_{1},q}_{n_{2}n_{3}}\left(p_{12}\right)
\\
=\frac{p_{1}^{n_{1}}}{\Gamma\left(n_{1}+1\right)}
\sum_{s=0}^{\infty}\frac{\left(-n_{2}\right)_{s}}{\left(n_{3}+s+1\right)!}m^{n_{2}+q-s}_{n_{3}+s+1}\left(p_{2}\right),
\end{multline}
\begin{multline}\label{eq:62_GALTERM}
m^{n_{1}}_{n_{2}}\left(p\right)
\\
=2^{n_{1}}U\left(n_{2}+1,n_{1}+n_{2}+2,p\right)\Gamma\left(n_{2}+1\right)e^{-p},
\end{multline}
and,
\begin{multline}\label{eq:63_GALTERK}
k^{n_{1},q}_{n_{2},n_{3}}\left(p_{12}\right)
\\
=\frac{p_{1}^{n_{1}}}{\Gamma\left(n_{1}+1 \right)}2^{n_{2}+n_{3}+1}B\left(n_{2}+1,n_{3}+1, \frac{1}{2}\right) 
\\
\times E_{-\left(n_{2}+n_{3}+q+1\right)}\left(p_{2}\right),
\end{multline}
with,
\begin{multline}\label{eq:64_ALTERNATIVE6}
U\left(a,b;z\right)
=\frac{\Gamma\left(b-1\right)}{\Gamma\left(a\right)}{_{1}F_{1}}\left(a-b+1,2-b;z\right) 
\\
+\frac{\Gamma\left(1-b\right)}{\Gamma\left(a-b+1\right)}{_{1}F_{1}}\left(a;b;z\right),
\end{multline}
are the tricomi confluent hyper$-$geometric functions with ${_{1}F_{1}}$ are the Kummer confluent hypergeometric function \cite{107_Abramowitz_1972, 108_Arfken_1985} and $B\left(a,b \right)$, $B\left(a,b, z\right)$ are the beta functions and incomplete beta functions, respectively \cite{109_Oldham_2009}.\\
if the parameter $p_{3}\neq0$;\\
\begin{multline}\label{eq:65_GALTER2}
{\mathcal{G}^{n_{1},q}_{n_{2}n_{3}}}(p_{123})
=\frac{p_{1}^{n_{1}}}{\Gamma\left(n_{1}+1 \right)}
\sum_{s=0}^{\infty}
\frac{p_{3}^{s}}{\Gamma\left(s+1 \right)}
\frac{1}{s+q+1}
\\
\times \left\lbrace
\left(-1\right)^{s}J_{n_{2}n_{3}}^{s+q,q}\left(p_{2}\right)
+\left(-1\right)^{q}J_{n_{3}n_{2}}^{s+q,q}\left(p_{2}\right)
\right\rbrace,
\end{multline}
where,
\begin{multline}\label{eq:66_GALTERJ}
J_{n_{1}n_{2}}^{s,q}\left(p\right)
=\left(\frac{s+1}{s}\right)
\\
\times \left\lbrace
J_{n_{1}n_{2}}^{s-1,q+1}\left(p\right)
-J_{n_{1}+1n_{2}}^{s-1,q}\left(p\right)
\right\rbrace
\end{multline}
\begin{align}\label{eq:67_GALTERJ0}
J_{n_{1}n_{2}}^{0,q}\left(p_{01}\right)
=k^{1,q}_{n_{1}n_{2}}\left(p_{01}\right)
-\frac{1}{2}
l^{1,q}_{n_{1}n_{2}}\left(p_{01}\right),
\end{align}
with, $p_{01}=\left\lbrace 1,p \right\rbrace$.\\
Explicit form of the $J^{s,q}$ functions involve Appell hyper$-$geometric function \cite{110_Appell_1925} and their are given as,
\begin{multline}\label{eq:68_GALTERJSINTEG}
J_{n_{1}n_{2}}^{s,q}\left(p\right)\\
=\int_{1}^{\infty}F_{1}\left(s+1;-n_{1},-n_{2};s+2;\frac{1}{\xi},-\frac{1}{\xi} \right)\\
\times \xi^{n_{1}+n_{2}+q}e^{-p\xi}.
\end{multline}
where, $F_{1}$ are the Appell functions,
\begin{multline}\label{eq:69_APPELLFUNC}
F_{1}\left(a;b_{1},b_{2};c;z_{1},z_{2} \right)=
\frac{\Gamma\left(c\right)}{\Gamma\left(a\right)\Gamma\left(a-c\right)}\\
\times \int_{0}^{1}u^{a-1}\left(1-u\right)^{c-a-1}
\left(1-uz_{1}\right)^{-b_{1}}
\left(1-uz_{2}\right)^{-b_{2}}du.
\end{multline}
\section{\label{sec:resdiscuss}Results and Discussions}
\begin{table*}[htp!]
\caption{\label{tab:OverlapCT} Convergence behavior of the analytical solution of two$-$center overlap integrals via Eqs. (\ref{eq:59_GALTER1}, \ref{eq:65_GALTER2}).}
\begin{ruledtabular}
\begin{tabular}{ccccccccc}
$n$ & $l$ & $n'$ & $l'$ & $\lambda$ & $\zeta$ & $\zeta'$ & $R$ & Results
\\
\hline
$5.1$ & $4$ & $5.1$ & $4$ & $0$ & $2.5$ & $2.5$ & $2.0$ & \begin{tabular}[c]{@{}l@{}l@{}l@{}l@{}l@{}l@{}l@{}}
\underline{3.68837 33855 08336 58641 31918 22868 35839 E-01}\footnotemark[1]\\
\underline{3.68837 33855 08336 58641 31918 22868 35839 E-01} (0100)\footnotemark[2]\\ 
\underline{3.68837 33855 08336 58641 31918 22868 35839 E-01} (0075)\footnotemark[2]\\ 
\underline{3.68837 33855 08336 58641 31918 2}3395 14728 E-01 (0050)\footnotemark[2]\\ 
\underline{3.68837 33855 0833}7 99988 61283 29326 92589 E-01 (0025)\footnotemark[2]\\ \\
\underline{3.68837 33855 0}5726 37942 01568 75075 77285 E-01 (1500)\footnotemark[3]\\ \underline{3.68837 33855 0}2829 31225 21439 34449 97437 E-01 (1250)\footnotemark[3]\\ \underline{3.68837 3385}4 94605 78092 61548 46231 82180 E-01 (1000)\footnotemark[3]\\ \underline{3.68837 3385}4 63771 70501 74331 34118 31567 E-01 (0750)\footnotemark[3]\\ \underline{3.68837 3385}2 74417 43376 93079 44890 38419 E-01 (0500)\footnotemark[3]\\ \underline{3.68837 338}15 49121 07703 85081 75542 56719 E-01 (0250)\footnotemark[3]\\ \underline{3.68837 3}2224 55592 65438 31561 52778 02193 E-01 (0100)\footnotemark[3]\\
\underline{3.68837} 07606 36279 99583 24709 21920 68306 E-01 (0050)\footnotemark[3]
\end{tabular}
\\
\\
$3.8$ & $0$ & $5.5$ & $0$ & $0$ & $2.31$ & $0.77$ & $2.0$ & \begin{tabular}[c]{@{}l@{}l@{}l@{}l@{}l@{}l@{}l@{}}
\underline{2.90802 04650 66341 47700 88166 91317 05703 E-01}\footnotemark[1]\\
\underline{2.90802 04650 66341 47700 88166 91317 05703 E-01} (0030)\footnotemark[4]\\ 
\underline{2.90802 04650 66341 47700 88166 9131}6 83635 E-01 (0025)\footnotemark[4]\\ 
\underline{2.90802 04650 66341 47700 881}35 45107 55970 E-01 (0020)\footnotemark[4]\\
\underline{2.90802 04650 66341} 38346 37860 92398 59679 E-01 (0015)\footnotemark[4]\\
\underline{2.90802 04649 604}01 30605 90542 27209 54872 E-01 (0010)\footnotemark[4]\\
\underline{2.90}792 57796 56773 38639 56179 86886 71667 E-01 (0005)\footnotemark[4]\\ \\
\underline{2.90802 04650 66341 47}698 70929 79988 77235 E-01 (1500)\footnotemark[3]\\ 
\underline{2.90802 04650 66341 47}677 90448 35045 93450 E-01 (1000)\footnotemark[3]\\ 
\underline{2.90802 04650 66341 47}578 14270 34479 60056 E-01 (0750)\footnotemark[3]\\ 
\underline{2.90802 04650 66341 4}6393 45932 16932 70044 E-01 (0500)\footnotemark[3]\\ 
\underline{2.90802 04650 66341 4}5280 68510 67242 29880 E-01 (0450)\footnotemark[3]\\ 
\underline{2.90802 04650 6634}0 71689 25970 04673 72415 E-01 (0250)\footnotemark[3]\\ 
\underline{2.90802 04650 66}165 26767 66810 48702 51777 E-01 (0100)\footnotemark[3]\\
\underline{2.90802 04650} 53956 39922 07646 27387 87619 E-01 (0050)\footnotemark[3]
\end{tabular}
\\
\footnotetext[1]{Ref. \cite{67_Bagci_2014}, benchmark result obtained via global-adaptive method with Gauss-Kronrod extension.}
\footnotetext[2]{Results obtained via Eq. (\ref{eq:59_GALTER1}).}
\footnotetext[3]{Ref. \cite{67_Bagci_2014}, results obtained via binomial expansion method.}
\footnotetext[4]{Results obtained via Eq. (\ref{eq:65_GALTER2}).}
\footnotetext[0]{The values in parenthesis are upper limit of summations.}
\end{tabular}
\end{ruledtabular}
\end{table*}
The difficulties associated with using the point$-$like model of nucleus in the four$-$component relativistic method is discussed. The results presented are obtained from solution of a generalized eigenvalue equation (Eq. (\ref{eq:8_DHFRFOURCOMPACT})). The single$-\zeta$ basis set approximation is used in a linear combination of Slater$-$type spinor orbital basis. Calculations are performed using a computer program written in the Mathematica programming language. Schur decomposition \cite{111_Moler_1973} and Powel optimization method \cite{112_Powell_1964} enabled us to obtain variationally optimum values for energy eigenvalues.

As a continuation to our previous results \cite{51_Bagci_2016} given for the hydrogen$-$like tin atom that prove clear separation between positive$-$ and negative$-$energy spectrum, in this study, the upper limit of summation in LCAS is increased while investigating the degenerate excited energy states. Fixed values for exponent $n$ in radial functions are defined. The difference $\left(\Delta E \right)$ between $E_{2s_{1/2}}, E_{2p_{1/2}}$ and $E_{3s_{1/2}}, E_{3p_{1/2}}$ energy states are plotted in Figure{\ref{fig:DIRACSTOEIGEN}}. It can be seen from this figure that the smallest value for $\Delta E$ is found when $n=\gamma^{-}=\sqrt{\kappa^2- \left(\alpha Z \right)^2}$ and the largest one when $n=\gamma^{+}=\sqrt{\kappa^2+ \left(\alpha Z \right)^2}$. This figure presents results multiplied by $E+03$, There is almost no difference between the results obtained for $n=\gamma^{-}$ and $n= \vert \kappa \vert$ while upper limit of summation $N$, $N=25$. Any value for $n$ thereof can be used to correctly represent a physical system. The exponent $n$ may even be used as a variational parameter. The choice however, depends on characteristics of a system. This becomes more apparent when calculating atoms with nuclear charge $Z$, $Z>137$. In this case variational stability is not guaranteed for all values of $n$. The most conspicuous example is to consider the radial exponent $n$ as $n=\gamma^{-}$. The energy eigenvalues obtained from the Dirac equation solution become imaginary. Real eigenvalues for any value of nuclear charge are obtained by considering $n$ independent from speed of light, or $n\geq \vert \kappa \vert$. The relationship between variational stability, determination of critical nuclear charge $Z_{c}$ and set of values for radial exponent $n$ in a basis are conserved.

This is shown in Table \ref{tab:HlikeECT} for nuclear charge $Z$, $110 \leq Z \leq 160$. Two sets of values with $n= \vert \kappa \vert + \epsilon$, $0< \epsilon < 1$ and $n= 2\vert\kappa \vert$ are defined for $n$. The upper limit of summation $N$ in linear combination of atomic spinors is determined as  $N=64$ which is mean highest limit principal quantum number is $n=8$. Note that this differs from the radial exponent $n$. The principal quantum numbers represent a sequence of electron configurations to be included to the linear combination. The ground and some excited states of hydrogen$-$like atoms depending on nuclear charge are given. It can be seen from this table that unlike considering the nucleus as finite$-$sized, the difference between degenerate energy states increases much more slowly. Some results obtained are consistent with those found in \cite{113_Pieper_1969} even when $Z=150$ or $Z=160$. Among other things, this table gives benchmark values which certainly need to be dealt with thoroughly. Increasing the upper limit of summation in LCAS, further investigation on electronic energy states, clarification of dependence between nuclear charge and radial exponent will be considered in future work. \\
A test calculation is performed for upper limit of summation $N$, $N=256$ (here, the highest limit of principal quantum number $n$ is $n=16$ and radial exponent $n$ is $n= \vert \kappa \vert$.) for nuclear charge $Z$, $Z=136, 137$ and $Z=138$. The ground state energies are obtained as $408.46 \hspace{1mm} KeV$, $422.17 \hspace{1mm} KeV$, $436.86 \hspace{1mm} KeV$, respectively. Variational stability is maintained even while such a quite large basis set is used. Another test calculation is performed for nuclear charge $Z$, $Z=140$ with upper limit of summation $N$, $N=100$. The radial exponent are taken to be $n= \vert \kappa \vert$ and $n= \gamma^{+}$. The ground state energies are found as $448.48 \hspace{1mm} KeV$, $425.40 \hspace{1mm} KeV$ (this value is compatible to that found in \cite{113_Pieper_1969}.), respectively. Note that absolute values for eigenvalues obtained from solution are given in this study. Finally, it is possible to conclude with the results presented in this study for hydrogen$-$like atoms that the so called "catastrophe" that previously emerged for a charge numbers $Z$ with $Z>137$, in  solving  the  Dirac  equation  with a potential corresponding to a point$-$charge no longer applies.

Additional mathematical difficulties arise in relativistic calculations of more complex systems such as molecules. One of the most challenging among them is pointed out in the above section \ref{sec:stso} and its solution given in section \ref{sec:molauxrev}. Results for two$-$center overlap integrals are presented in Table \ref{tab:OverlapCT}, accordingly. An infinite series expansion occurs in Eq. (\ref{eq:65_GALTER2}) while $p_{3} \neq 0$. This results from series expansion of exponential functions $e^{z}$, $z=-p3\nu$. The exponential function is uniformly convergent for the entire complex plane for any $z$ with $z<\infty$. Convergence behavior of relativistic molecular auxiliary functions given Eq. (\ref{eq:54_AUXG}) through Eq. ({\ref{eq:50_AUX}}) and Eq. (\ref{eq:65_GALTER2}) are tested in this table. From the earlier version \cite{73_Bagci_2017} of the present paper were used by the author to derive fully analytical formulae \cite{71_Bagci_2018_1, 72_Bagci_2018_2} for Eq. ({\ref{eq:66_GALTERJ}). Here, this equation is numerically calculated since the convergence properties of Eq. (\ref{eq:65_GALTER2}) should be investigated initially. The results in Table \ref{tab:OverlapCT} (dotted lines) for some values of principal quantum numbers and orbital parameters shows that this task has been accomplished.


\begin{thebibliography}{plainnat}

\bibitem{1_Hartree_1928} D. R. Hartree, Math. Proc. Camb. Philos. Soc. {\bf 24}(1), 89 (1928). doi: \url{https://doi.org/10.1017/S0305004100011919}.

\bibitem{2_Fock_1930} V. A. Fock, Zeitschrift f{\"u}r Physik \textbf{62}(11-12), 795 (1930). doi: \url{https://doi.org/10.1007/BF01330439}.

\bibitem{3_Fock_1978} V. A. Fock, {\sl Fundamentals of Quantum Mechanics} (MIR Publishers, Moscow, 1978).

\bibitem{4_Landau_1982} V. B. Berestetskii, E. M. Lifshitz and L. P. Pitaevskii, {\sl Quantum Electrodynamics, Landau and Lifshitz Course of Theoretical Physics} Volume 4, 2nd edition (Pergamon, London 1982).

\bibitem{5_Lindgren_2011} I. Lindgren, {Relativistic many$-$body theory: a new field$-$theoretical approach} (Springer, New York, 2011).

\bibitem{6_Ford_1974} B. Ford and G. Hall, Computer Phys. Commun. \textbf{8}(5), 337 (1974). doi: \url{https://doi.org/10.1016/0010-4655(74)90011-3}.

\bibitem{7_Roothaan_1951} C. C. J. Roothaan, Rev. Mod. Phys. \textbf{23}(2), 69 (1951). doi: \url{https://link.aps.org/doi/10.1103/RevModPhys.23.69}.

\bibitem{8_Grant_1961} I. P. Grant, Proc. Roc. Soc. Lond. A \textbf{262}, 555 (1961). doi: \url{https://doi.org/10.1098/rspa.1961.0139}.

\bibitem{9_Grant_1965} I. P. Grant, Proc. Phys. Soc. Lond. \textbf{86}(3), 523 (1965). doi: \url{https://doi.org/10.1088%2F0370-1328%2F86%2F3%2F311}.

\bibitem{10_Kim_1967} Yong-Ki Kim, Phys. Rev. \textbf{154}(1), 17 (1967). doi: \url{https://link.aps.org/doi/10.1103/PhysRev.154.17}.

\bibitem{11_Leclercq_1970} J. M. Leclercq, Phys. Rev. A {\bf 1}(5), 1358 (1970). doi: \url{https://link.aps.org/doi/10.1103/PhysRevA.1.1358}.

\bibitem{12_Laaksonen_1988} L. Laaksonen, I. P. Grant and S Wilson, J. Phys. B: At. Mol. Opt. Phys. \textbf{21}(11), 1969 (1988). doi: \url{https://doi.org/10.1088%2F0953-4075%2F21%2F11%2F013}

\bibitem{13_Quiney_1987} H. M. Quiney, I. P. Grant and S. Wilson, J. Phys. B: At. Mol. Phys. {\bf 20}(7), 1413 (1987). doi: \url{https://doi.org/10.1088/0022-3700/20/7/010}.

\bibitem{14_Malli_1975} G. Malli and J. Oreg, J. Chem. Phys. \textbf{63}(2), 830 (1975). doi: \url{https://doi.org/10.1063/1.431364}.

\bibitem{15_Matsuoka_1980} O. Matsuoka, N. Suzuki, T. Aoyama and G. Malli, J. Chem. Phys. \textbf{73}(3), 1320 (1980). doi: \url{https://doi.org/10.1063/1.440245}.

\bibitem{16_Pisani_1994} L. Pisani and E. Clementi, J. Comput. Chem. \textbf{15}(4), 466 (1994). doi: \url{https://onlinelibrary.wiley.com/doi/abs/10.1002/jcc.540150410}.

\bibitem{17_Yanai_2001} T. Yanai, T. Nakajima, Y. Ishikawa and K. Hirao, J. Chem. Phys. \textbf{114}(15), 6526 (2001). doi: \url{https://doi.org/10.1063/1.1356012}.

\bibitem{18_Quiney_2002} H. M. Quiney, P. Belanzoni and A. Sgamellotti, Theor. Chem. Acc. {\bf 108}(2), 113 (2002). doi: \url{https://doi.org/10.1007/s00214-002-0369-3}.

\bibitem{19_Belpassi_2008} L. Belpassi, F. Tarantelli, A. Sgamellotti and H. M. Quiney, Phys. Rev. B {\bf 77}(23), 233403 (2008). doi: \url{https://link.aps.org/doi/10.1103/PhysRevB.77.233403}.

\bibitem{20_Sucher_1980} J. Sucher, Phys. Rev. A {\bf 22}(2), 348 (1980). doi: \url{https://link.aps.org/doi/10.1103/PhysRevA.22.348}.

\bibitem{21_Grant_2007} I. P. Grant, {\sl Relativistic Quantum Theory of Atoms and Molecules} (Springer,  New York, 2007).

\bibitem{22_Dirac_1930} P. A. M. Dirac, {\sl The principles of quantum mechanics} (Oxford Science Publications, Oxford, 1930).

\bibitem{23_Foldy_1950} L. L. Foldy and S. A. Wouthuysen, Phys. Rev. {\bf 78}(1), 29 (1950). doi: \url{https://link.aps.org/doi/10.1103/PhysRev.78.29}.

\bibitem{24_Greiner_2000} W. Greiner, {\sl Relativistic Quantum Mechanics: Wave Equations} (Springer, Berlin, 2000).

\bibitem{25_Grant_2010} I. P. Grant, J. Phys. B {\bf 43}(7), 074033 (2010). doi: \url{https://doi.org/10.1088%2F0953-4075%2F43%2F7%2F074033}.

\bibitem{26_Grant_1980} I. P. Grant, B.J.McKenzie, P. H. Norrington, D. F. Mayers and N. C. Pyper, Comput. Phys. Commun. {\bf 21}(2), 207 (1980). doi: \url{https://doi.org/10.1016/0010-4655(80)90041-7}.

\bibitem{27_Ishikawa_1991} Y. Ishikawa, H. M. Quiney and G. L. Malli, Phys. Rev. A {\bf 43}(7), 3270 (1991). doi: \url{https://link.aps.org/doi/10.1103/PhysRevA.43.3270}.

\bibitem{28_Ishikawa_1993} Y. Ishikawa and H. M. Quiney, Phys. Rev A {\bf 47}(3), 1732 (1993). doi: \url{https://link.aps.org/doi/10.1103/PhysRevA.47.1732}.

\bibitem{29_Koc_1994} K. Koc and Y. Ishikawa, Phys. Rev A {\bf 49}(2), 794 (1994). doi: \url{https://link.aps.org/doi/10.1103/PhysRevA.49.794}.

\bibitem{30_Ishikawa_1997} Y. Ishikawa, K. Koc and W. H. E. Schwarz, Chem. Phys. {\bf 225}(1), 239 (1997). doi: \url{https://doi.org/10.1016/S0301-0104(97)00267-X}.

\bibitem{31_Ishikawa_2001} Y. Ishikawa and M. J. Vilkas, J. Mol. Struct (Theocem) {\bf 573}(1), 139 (2001). doi: \url{https://doi.org/10.1016/S0166-1280(01)00540-1}.

\bibitem{32_Quiney_1997} H. M. Quiney, H. Skaane and I. P. Grant, J. Phys. B: At. Mol. Opt. Phys. {\bf 30}(23), L829 (1997). doi: \url{https://doi.org/10.1088%2F0953-4075%2F30%2F23%2F001}.

\bibitem{33_Quiney_1998} H. M. Quiney, H. Skaane and I. P. Grant, Adv. Quant. Chem. {\bf 32}, 1 (1998). doi: \url{https://doi.org/10.1016/S0065-3276(08)60405-0}.

\bibitem{34_Quiney_2004} H. M. Quiney, V. N. Glushkov and S. Wilson, Int. J. Quant. Chem. {\bf 99}(6), 950 (2004). doi: \url{https://onlinelibrary.wiley.com/doi/abs/10.1002/qua.20146}.

\bibitem{35_Saue_1996} T. Saue, K. Faegri and O. Gropen, Chem. Phys. Lett. {\bf 263}(3), 360 (1996). doi: \url{https://doi.org/10.1016/S0009-2614(96)01250-X}.

\bibitem{36_Saue_1997} T. Saue, K. Faegri, T. Helgaker and O. Gropen, Mol. Phys. {\bf 91}(5), 937 (1997). doi: \url{https://www.tandfonline.com/doi/abs/10.1080/002689797171058}.

\bibitem{37_Saue_2011} T. Saue, ChemPhysChem {\bf 12}(17), 3077 (2011). doi: \url{https://onlinelibrary.wiley.com/doi/abs/10.1002/cphc.201100682}.

\bibitem{38_Pyykko_2012} P. Pyykk{\"o}, Annu. Rev. Phys. Chem. {\bf 63}, 45 (2012). doi: \url{https://doi.org/10.1146/annurev-physchem-032511-143755}.

\bibitem{39_Motoumba_2019} E. B. Motoumba, S. E. Yoca, P. Palmeri and P. Quinet, Journal of Quantitative Spectroscopy and Radiative Transfer {\bf 227}, 130 (2019). doi: \url{https://doi.org/10.1016/j.jqsrt.2019.01.028}.

\bibitem{40_Si_2018} R. Si, X. L. Guo, T. Brage, C. Y. Chen, R. Hutton and C. F. Fischer, Phys. Rev. A {\bf 98}(1), 012504 (2018). doi: \url{https://link.aps.org/doi/10.1103/PhysRevA.98.012504}.

\bibitem{41_Indelicato_1995} P. Indelicato, Phys. Rev. A {\bf 51}(2), 1132 (1995). doi: \url{https://link.aps.org/doi/10.1103/PhysRevA.51.1132}.

\bibitem{42_Kutzelnigg_2012} W. Kutzelnigg, Chem. Phys. {\bf 395}, 16 (2012). doi: \url{https://doi.org/10.1016/j.chemphys.2011.06.001}.

\bibitem{43_Fleig_2012} T. Fleig, Chem. Phys. {\bf 395}, 2 (2012). doi: \url{https://doi.org/10.1016/j.chemphys.2011.06.032}.

\bibitem{44_Schwarz_1982} W.H. E. Schwarz and H.Wallmeier, Mol. Phys. {\bf 46}(5), 1045 (1982). doi: \url{https://doi.org/10.1080/00268978200101771}.

\bibitem{45_Schwarz_1982} W. H. E. Schwarz and E. Wechsel-Trakowski, Chem. Phys. Lett. {\bf 85}(1), 94 (1982). doi: \url{https://doi.org/10.1016/0009-2614(82)83468-4}.

\bibitem{46_Drake_1981} G. W. F. Drake and S. P. Goldman, Phys. Rev A {\bf 23}(5), 2093 (1981). doi: \url{https://link.aps.org/doi/10.1103/PhysRevA.23.2093}.

\bibitem{47_Goldman_1985} S. P. Goldman, Phys. Rev. A {\bf 31}(6), 3541 (1985). doi: \url{https://link.aps.org/doi/10.1103/PhysRevA.31.3541}.

\bibitem{48_Lee_1982} Y. S. Lee and A. P. McLean, J. Chem. Phys. {\bf 76}(1), 735 (1982). doi: \url{https://doi.org/10.1063/1.442680}.

\bibitem{49_Stanton_1984} R. E. Stanton and S. Havriliak, J. Chem. Phys. {\bf 81}(4), 1910 (1984). doi: \url{https://doi.org/10.1063/1.447865}.

\bibitem{50_Dyall_2012} K. G. Dyall, Chem. Phys. {\bf 395}, 35 (2012). doi: \url{https://doi.org/10.1016/j.chemphys.2011.07.009}.

\bibitem{51_Bagci_2016} A Ba{\u g}c{\i}, P. E. Hoggan, Phys. Rev E {\bf 94}(1), 013302 (2016). doi: \url{https://link.aps.org/doi/10.1103/PhysRevE.94.013302}.

\bibitem{52_Davydov_1976} A. S. Davydov, {\sl Quantum Mechanics} (Pergamon, London, 1976).

\bibitem{53_Wigner_1959} E. P. Wigner, {\sl Group Theory and its Application to the Quantum Mechanics of Atomic Spectra} (Academic Press, New York, 1959).

\bibitem{54_Varshalovich_1988} D. A. Varshalovich, A. N. Moskalev  and V. K. Khersonski, {\sl Quantum  theory  of  angular  momentum} (World Scientific,Singapore, 1988)

\bibitem{55_Wei_1999} L. Wei, Comput. Phys. Commun. {\bf 120}(2), 222 (1999). doi: \url{https://doi.org/10.1016/S0010-4655(99)00232-5}.

\bibitem{56_Condon_1970} E. U. Condon and G. H. Shortley, {\sl The Theory of Atomic Spectra} (Cambridge University Press, Cambridge, 1970).

\bibitem{57_Kato_1957} T. Kato, Commun. Pure. Appl. Math. {\bf 10}(2), 151 (1957). doi: \url{https://onlinelibrary.wiley.com/doi/abs/10.1002/cpa.3160100201}.

\bibitem{58_Agmon_1985} S. Agmon, Lect. Notes Math. {\bf 1159}, 1 (1985). doi: \url{https://doi.org/10.1007/BFb0080331}.

\bibitem{59_Singulariy_2017} J. Karwowski, {\sl Dirac Operator and Its Properties}, pp. 3-49; D. Andrare, {\sl Nuclear Charge Density and Magnetization Distributions}, pp. 51-82; K. G. Dyall, {\sl One-Particle Basis Sets for Relativistic Calculations}, pp. 83-106; C. W{\"u}llen, {\sl Relativistic Self-Consistent Fields}, pp. 107-127; S. Shao, Z. Li, W. Liu, {\sl Coalescence Conditions of Relativistic Wave Functions}, pp. 497-530 in Handbook of Relativistic Quantum Chemistry, edited by W. Liu (Springer-Verlag, Berlin, 2017).

\bibitem{60_Mourou_2006} G. A. Mourou, T. Tajima, and S. V. Bulanov, Rev. Mod. Phys. {\bf 78}(2), 309 (2006). doi: \url{https://link.aps.org/doi/10.1103/RevModPhys.78.309}.

\bibitem{61_Selsto_2009} S. Selsto, E. Lindroth, and J.Bengtsson, Phys. Rev.A {\bf 79}(4), 043418 (2009). doi: \url{https://link.aps.org/doi/10.1103/PhysRevA.79.043418}.

\bibitem{62_Pohl_2013} R. Pohl, R. Gilman, G. A. Miller and K. Pachucki, Annu. Rev. Nucl. Part. Sci. {\bf 63}, 175 (2013). doi: \url{https://doi.org/10.1146/annurev-nucl-102212-170627}.

\bibitem{63_Jungmann_1992} K. Jungmann, Z. Physik C Particles and Fields {\bf 56}(1), S59 (1992). doi: \url{https://doi.org/10.1007/BF02426776}.

\bibitem{64_Olver_2018} P. J. Olver, {\sl Complex analysis and conformal mapping}, lecture notes, http://www-users.math.umn.edu/$\sim$olver/ln$\textunderscore$/cml.pdf, August 15, 2018.

\bibitem{65_Weniger_2008} E. J. Weniger, J. Phys. A: Math. Theor. {\bf 41}(42), 425207 (2008). doi: \url{https://doi.org/10.1088%2F1751-8113%2F41%2F42%2F425207}.

\bibitem{66_Weniger_2012} E. J. Weniger, J. Math. Chem. {\bf 50}(1), 17 (2012). doi: \url{https://doi.org/10.1007/s10910-011-9914-4}.

\bibitem{67_Bagci_2014} A. Ba{\u{g}}c{\i} and P. E. Hoggan, Phys. Rev. E \textbf{89}(5), 053307 (2014). doi: \url{https://link.aps.org/doi/10.1103/PhysRevE.89.053307}.

\bibitem{68_Bagci_2015} A. Ba{\u{g}}c{\i} and P. E. Hoggan, Phys. Rev. E, \textbf{91}(2), 023303 (2015). doi: \url{https://link.aps.org/doi/10.1103/PhysRevE.91.023303}.

\bibitem{69_Bagci_2015} A. Ba{\u{g}}c{\i} and P. E. Hoggan, Phys. Rev. E, \textbf{92}(4), 043301 (2015). doi: \url{https://link.aps.org/doi/10.1103/PhysRevE.92.043301}.

\bibitem{70_MathematicaProg} \url{http://www.wolfram.com/mathematica}.

\bibitem{71_Bagci_2018_1} A. Ba{\u{g}}c{\i} and P. E. Hoggan, Rendiconti Lincei. Scienze Fisiche e Naturali {\bf 29}(1), 191 (2018). doi: \url{https://doi.org/10.1007/s12210-018-0669-8}.

\bibitem{72_Bagci_2018_2} A. Ba{\u{g}}c{\i}, P. E. Hoggan and M. Adak, Rendiconti Lincei. Scienze Fisiche e Naturali {\bf 29}(4), 765 (2018). doi: \url{https://doi.org/10.1007/s12210-018-0734-3}.

\bibitem{73_Bagci_2017} A. Ba{\u{g}}c{\i}, {\sl Notes on mathematical difficulties arising in relativistic SCF approximation}, arXiv :1603.02307 [physics.chem-ph] (2017).

\bibitem{74_Thaller_1992} B. Thaller, {\sl The Dirac Equation} (Springer-Verlag, Amsterdam, 1992).

\bibitem{75_Autschbach_2000} J. Autschbach and W. H. E. Schwarz, Theor. Chem. Acc. {\bf 104}(1), 82 (2000). doi: \url{https://doi.org/10.1007/s002149900108}.

\bibitem{76_Schwerdtfeger_2015} P. Schwerdtfeger, L. F. Pa{\u s}teka, A. Punnett and P. O. Bowman, Nuclear Phys. A {\bf 944}, 551 (2015). doi: \url{https://doi.org/10.1016/j.nuclphysa.2015.02.005}.

\bibitem{77_Brown_1951} G. E. Brown and G. D. Ravenhall, Proc. Roy. Soc. London A {\bf 208}(1095), 552(1951). doi: \url{https://royalsocietypublishing.org/doi/abs/10.1098/rspa.1951.0181}.

\bibitem{78_Dyall_1989} K. G. Dyall, I. P. Grant, C. T. Johnson, F. A. Parpia and E. P. Plummer, Comput. Phys. Commun. {\bf 55}(3), 425 (1989). doi: \url{https://doi.org/10.1016/0010-4655(89)90136-7}.

\bibitem{79_Visscher_1994} L. Visscher, O. Visser, P. J. C. Aerts, H. Merenga and W. C. Nieuwpoort, Comput. Phys. Commun. {\bf 81}(1), 120 (1994). doi: \url{https://doi.org/10.1016/0010-4655(94)90115-5}.

\bibitem{80_DIRAC_2018} DIRAC, {\sl a relativistic ab initio electronic structure program}, Release DIRAC18 (2018),  written by T. Saue, L. Visscher, H. J. {\relax Aa}. Jensen, and R. Bast, with contributions from V. Bakken, K. G. Dyall, S. Dubillard, U. Ekstr{\"o}m, E. Eliav, T. Enevoldsen, E. Fa{\ss}hauer, T. Fleig, O. Fossgaard, A. S. P. Gomes, E. D. Hedeg{\aa}rd, T. Helgaker, J. Henriksson, M. Ilia{\v{s}}, Ch. R. Jacob, S. Knecht, S. Komorovsk{\'y}, O. Kullie, J. K. L{\ae}rdahl, C. V. Larsen, Y. S. Lee, H. S. Nataraj, M. K.Nayak, P. Norman, G. Olejniczak, J. Olsen, J. M. H. Olsen, Y. C. Park, J. K. Pedersen, M. Pernpointner, R. di Remigio, K. Ruud, P. Sa{\l}ek, B. Schimmelpfennig, A. Shee, J. Sikkema, A. J. Thorvaldsen, J. Thyssen, J. van Stralen, S. Villaume, O. Visser, T. Winther, and S. Yamamoto (available at \url{https://doi.org/10.5281/zenodo.2253986} see also \url{http://www.diracprogram.org}).

\bibitem{81_Grant_2000} I. P. Grant and H. M. Quiney, Int. J. Quant. Chem. {\bf 80}(3), 283 (2000). doi: \url{https://onlinelibrary.wiley.com/doi/abs/10.1002/1097-461X%282000%2980%3A3%3C283%3A%3AAID-QUA2%3E3.0.CO%3B2-L}.

\bibitem{82_Shiozaki_2018} T. Shiozaki, Wires Comput. Mol. Sci. {\bf 8}(1), e1331 (2018). doi: \url{https://onlinelibrary.wiley.com/doi/abs/10.1002/wcms.1331}.

\bibitem{83_Hofmann_2011} S. Hofmann, Radiochim Acta, {\bf 99}(7-8), 405 (2011). doi: \url{https://doi.org/10.1524/ract.2011.1854}.

\bibitem{84_Grainer_2015} W. Grainer, {\sl Dedication to Prof. Mikhail Itkis}, pp.1-8; G. M{\"u}nzenberg, {\sl SHE Research at GSI$-$Historical Remarks and New Ideas}, pp.9-20 in Nuclear Physics Present and Future, edited by W. Greiner (Springer-Verlag, New York, 2015).

\bibitem{85_Oganessian_2011} Yu. Ts. Oganessian, Radiochim Acta {\bf 99}(7-8), 429 (2011). doi: \url{https://doi.org/10.1524/ract.2011.1860}.

\bibitem{86_Manjunatha_2019} H. C. Manjunatha, K. N. Sridhar and H. B. Ramalingam, Nuc. Phys. A {\bf 981}, 17 (2019). doi: \url{https://doi.org/10.1016/j.nuclphysa.2018.10.084}.

\bibitem{87_Slater_1930} J. C. Slater, Phys. Rev. {\bf 36}(1), 57 (1930). doi: \url{https://link.aps.org/doi/10.1103/PhysRev.36.57}.

\bibitem{88_Guseinov_1985} I. I. Guseinov, Phys. Rev. A {\bf 31}(5) 2851 (1985). doi: \url{https://link.aps.org/doi/10.1103/PhysRevA.31.2851}.

\bibitem{89_Lowdin_1950} P. L{\"o}wdin, J. Chem. Phys. {\bf 18}(3), 365 (1950). doi: \url{https://doi.org/10.1063/1.1747632}.

\bibitem{90_Press_1992} W H. Press, B. P. Flannery, S. A. Teukolsky, W. T. Vetterling, {\sl Cholesky Decomposition in Numerical Recipes in FORTRAN: The Art of Scientific Computing}, 2nd ed. (Cambridge University Press, Cambridge, England, 1992) pp. 89-91.

\bibitem{91_Schur_1909} I. Schur, Mathematische Annalen {\bf 66}(4), 488 (1909). doi: \url{https://doi.org/10.1007/BF01450045}.

\bibitem{92_Guseinov_1970} I. I. Guseinov, J. Phys. B: At. Mol. Phys. {\bf 3}(11), 1399 (1970). doi: \url{https://doi.org/10.1088%2F0022-3700%2F3%2F11%2F001}.

\bibitem{93_Chaudhry_2002} M. A. Chaudhry and S. M. Zubair, {\sl On a Class of Incomplete Gamma Functions with Applications}, (Chapman \& Hall/CRC, 2002)

\bibitem{94_Gautschi_2003} W. Gautschi, F. E. Harris  and N. M. Temme, Appl. Math. Lett. {\bf 16}(7), 1095 (2003). doi: \url{https://doi.org/10.1016/S0893-9659(03)90100-5}.

\bibitem{95_Blahak_2010} U. Blahak, Geoscientific Model Development, {\bf 3}(2), 329 (2010). doi: \url{https://www.geosci-model-dev.net/3/329/2010/}.

\bibitem{96_Gautschi_1979} W. Gautschi, ACM Trans. Math. Soft, {\bf 5}(4), 466 (1979). doi: \url{http://doi.acm.org/10.1145/355853.355863}.

\bibitem{97_Temme_1994_1} N. M. Temme, {\sl Computational Aspects of Incomplete Gamma Functions with Large Complex Parameters} in {\sl Approximation and Computation: A Festschrift in Honor of Walter Gautschi}, edited by Zahar R. V. M. (Birkh{\"a}user), (Springer, Heidelberg, Berlin, 1993) pp. 551-562.

\bibitem{98_Temme_1994_2} N. M. Temme, Probability in the Engineering and Informational Sciences, {\bf 8}(2), 291 (1994). doi: \url{https://doi.org/10.1017/S0269964800003417}.

\bibitem{99_Gil_2012} A. Gil, J. Segura and N. M. Temme, SIAM J. Sci. Comput, {\bf 34}(6), A2965 (2012). doi: \url{https://doi.org/10.1137/120872553}.

\bibitem{100_Weatherford_2005} C. A. Weatherford, E. Red, P. E. Hoggan, Mol. Phys. {\bf 103}(15-16) 2169 (2005). doi: \url{https://doi.org/10.1080/00268970500137261}.

\bibitem{101_Absi_2006} N. Absi and P. E. Hoggan, Int. J. Quant. Chem. {\bf 106}(14) 2881 (2006). doi: \url{https://onlinelibrary.wiley.com/doi/abs/10.1002/qua.21114}.

\bibitem{102_BoostCpp} \url{http://www.boost.org}.

\bibitem{103_Bezanson_2017} J. Bezanson, A. Edelman, S. Karpinski and V. B. Shah, Siam Rev. {\bf 59}(1), 65 (2017). doi: \url{https://doi.org/10.1137/141000671}.

\bibitem{104_Fieker_2017} C. Fieker, W. Hart, T. Hofmann and F. Johansson {\sl Nemo/Hecke: Computer Algebra and Number Theory Packages for the Julia Programming Language} in Proceedings of ISSAC '17 (New York, ACM, 2017) pp. 157-164.

\bibitem{105_Pople_1970} J. A. Pople and D. L. Beveridge, {\sl Approximate Molecular Orbital Theory}, (Mc-Graw Hill, New York, 1970).

\bibitem{106_Guseinov_2002} I. I. Guseinov and B. A. Mamedov, J. Mol. Mod. {\bf 8}(9), 272 (2002). doi: \url{https://doi.org/10.1007/s00894-002-0098-5}.

\bibitem{107_Abramowitz_1972} M. Abramowitz and I. A. Stegun, {\sl Handbook of Mathematical Functions with Formulas, Graphs, and Mathematical Tables} (New York, Dover, 1972).

\bibitem{108_Arfken_1985} G. Arfken, \textit{Mathematical Methods for Physicists} (Academic Press, 1985).

\bibitem{109_Oldham_2009} K. B. Oldham, J. C. Myland and J. Spanier, {\sl An Atlas of Functions}, (Springer-Verlag, New York, 2009).

\bibitem{110_Appell_1925} P. Appell {\sl Sur les fonctions hyperg{\'e}om{\'e}triques de plusieurs variables, les polyn{\^o}mes d'Hermite et autres fonctions sph{\'e}riques dans l'hyperespace}. M{\'e}morial des sciences math{\'e}matiques (Gauthier-Villars, Paris, France, 1925).

\bibitem{111_Moler_1973} C.B. Moler and G. W. Stewart, SIAM J. Numer. Anal. {\bf 10}(2), 241 (1973). doi: \url{https://doi.org/10.1137/0710024}.

\bibitem{112_Powell_1964} M. J. D. Powell, The Computer Journal, {\bf 7}(2), 155 (1964). doi: \url{https://doi.org/10.1093/comjnl/7.2.155}.

\bibitem{113_Pieper_1969} W. Pieper, W. Greiner, Z. Physik, {\bf 218}(4), 327 (1969). doi: \url{https://doi.org/10.1007/BF01670014}.



\end{thebibliography}
\end{document}